\documentclass[twocolumn,showpacs,aps,prc,amsmath,amssymb,superscriptaddress,groupeaddress]{revtex4}

\usepackage{graphicx}
\usepackage{bm}

\begin{document}

\preprint{APS/123-QED}

\title{$\beta$-decay of nuclei around $^{90}$Se. Search for signatures of a $N$=56 sub-shell closure relevant the r-process}

\author{M.~Quinn}
\affiliation{Institute of Structure and Nuclear Astrophysics, University of Notre Dame, Notre Dame, IN, USA}
\affiliation{Joint Institute for Nuclear Astrophysics, University of Notre Dame, Notre Dame, IN, USA}
\author{A.~Aprahamian}
\affiliation{Institute of Structure and Nuclear Astrophysics, University of Notre Dame, Notre Dame, IN, USA}
\affiliation{Joint Institute for Nuclear Astrophysics, University of Notre Dame, Notre Dame, IN, USA}
\author{J.~Pereira}
\email{pereira@nscl.msu.edu}
\affiliation{National Superconducting Cyclotron Laboratory, Michigan State University, E.~Lansing, MI, USA}
\affiliation{Joint Institute for Nuclear Astrophysics, Michigan State University, E.~Lansing, MI, USA}
\author{R.~Surman}
\affiliation{Joint Institute for Nuclear Astrophysics, University of Notre Dame, Notre Dame, IN, USA}
\affiliation{Department of Physics and Astronomy, Union College, Schenectady, NY, USA}
\author{O.~Arndt}
\affiliation{Institut f\"{u}r Kernchemie, Universit\"{a}t Mainz, Mainz, Germany}
\affiliation{Virtuelles Institut f\"{u}r Struktur der Kerne and Nuklearer Astrophysik, Mainz,
Germany}
\author{T.~Baumann}
\affiliation{National Superconducting Cyclotron Laboratory, Michigan State University, E.~Lansing, MI, USA}
\author{A.~Becerril}
\affiliation{National Superconducting Cyclotron Laboratory, Michigan State University, E.~Lansing, MI, USA}
\affiliation{Joint Institute for Nuclear Astrophysics, Michigan State University, E.~Lansing, MI, USA}
\affiliation{Department of Physics and Astronomy, Michigan State University, E.~Lansing, MI, USA}
\author{T.~Elliot}
\affiliation{National Superconducting Cyclotron Laboratory, Michigan State University, E.~Lansing, MI, USA}
\affiliation{Joint Institute for Nuclear Astrophysics, Michigan State University, E.~Lansing, MI, USA}
\affiliation{Department of Physics and Astronomy, Michigan State University, E.~Lansing, MI, USA}
\author{A.~Estrade}
\affiliation{National Superconducting Cyclotron Laboratory, Michigan State University, E.~Lansing, MI, USA}
\affiliation{Joint Institute for Nuclear Astrophysics, Michigan State University, E.~Lansing, MI, USA}
\affiliation{Department of Physics and Astronomy, Michigan State University, E.~Lansing, MI,
USA}
\author{D.~Galaviz}
\affiliation{National Superconducting Cyclotron Laboratory, Michigan State University, E.~Lansing, MI, USA}
\affiliation{Joint Institute for Nuclear Astrophysics, Michigan State University, E.~Lansing, MI, USA}
\author{T.~Ginter}
\affiliation{National Superconducting Cyclotron Laboratory, Michigan State University, E.~Lansing, MI, USA}
\author{M.~Hausmann}
\affiliation{Facility for Rare Isotope Beams, Michigan State University, E.~Lansing, MI, USA}\author{S.~Hennrich}
\affiliation{Institut f\"{u}r Kernchemie, Universit\"{a}t Mainz, Mainz, Germany} \affiliation{Joint Institute for Nuclear Astrophysics, Michigan State University,
E.~Lansing, MI, USA}
\affiliation{Virtuelles Institut f\"{u}r Struktur der Kerne and Nuklearer Astrophysik, Mainz, Germany}
\author{R.~Kessler}
\affiliation{Institut f\"{u}r Kernchemie, Universit\"{a}t Mainz, Mainz, Germany} \affiliation{Joint Institute for Nuclear Astrophysics, Michigan State University, E.~Lansing, MI, USA}
\affiliation{Virtuelles Institut f\"{u}r Struktur der Kerne and Nuklearer Astrophysik, Mainz, Germany}
\author{K.-L.~Kratz}
\affiliation{Virtuelles Institut f\"{u}r Struktur der Kerne and Nuklearer Astrophysik, Mainz, Germany}
\affiliation{Max Planck Institut f\"{u}r Chemie, Otto-Hahn-Institut, Mainz, Germany}
\author{G.~Lorusso}
\affiliation{National Superconducting Cyclotron Laboratory, Michigan State University, E.~Lansing, MI, USA}
\affiliation{Joint Institute for Nuclear Astrophysics, Michigan State University, E.~Lansing, MI, USA}
\affiliation{Department of Physics and Astronomy, Michigan State University, E.~Lansing, MI, USA}
\author{P.~F.~Mantica}
\affiliation{National Superconducting Cyclotron Laboratory, Michigan State University, E.~Lansing, MI, USA}
\affiliation{Department of Chemistry, Michigan State University, E.~Lansing, MI, USA}
\author{M.~Matos}
\affiliation{National Superconducting Cyclotron Laboratory, Michigan State University, E.~Lansing, MI, USA}
\affiliation{Joint Institute for Nuclear Astrophysics, Michigan State University, E.~Lansing, MI, USA}
\author{F.~Montes}
\affiliation{National Superconducting Cyclotron Laboratory, Michigan State University, E.~Lansing, MI, USA}
\affiliation{Joint Institute for Nuclear Astrophysics, Michigan State University, E.~Lansing,
MI, USA}
\author{B.~Pfeiffer}
\affiliation{Institut f\"{u}r Kernchemie, Universit\"{a}t Mainz, Mainz, Germany} \affiliation{Virtuelles Institut f\"{u}r Struktur der Kerne and Nuklearer Astrophysik, Mainz, Germany}
\author{M.~Portillo}
\affiliation{Facility for Rare Isotope Beams, Michigan State University, E.~Lansing, MI, USA}\author{S.~Hennrich}
\author{H.~Schatz}
\affiliation{National Superconducting Cyclotron Laboratory, Michigan State University, E.~Lansing, MI, USA}
\affiliation{Joint Institute for Nuclear Astrophysics, Michigan State University, E.~Lansing, MI, USA}
\affiliation{Department of Physics and Astronomy, Michigan State University, E.~Lansing, MI, USA}
\author{F.~Schertz}
\affiliation{Institut f\"{u}r Kernchemie, Universit\"{a}t Mainz, Mainz, Germany}
\affiliation{Joint Institute for Nuclear Astrophysics, Michigan State University, E.~Lansing,
MI, USA}
\affiliation{Virtuelles Institut f\"{u}r Struktur der Kerne and Nuklearer Astrophysik, Mainz, Germany}
\author{L.~Schnorrenberger}
\affiliation{Joint Institute for Nuclear Astrophysics, Michigan State University, E.~Lansing, MI, USA}
\affiliation{Institut f\"{u}r Kernphysik, TU Darmstadt, Darmstadt, Germany}
\author{E.~Smith} \affiliation{Joint Institute for Nuclear Astrophysics, Michigan State University, E.~Lansing, MI, USA}
\affiliation{Department of Physics, Ohio State University, Columbus, OH, USA}
\author{A.~Stolz} \affiliation{National Superconducting Cyclotron Laboratory, Michigan State University, E.~Lansing, MI, USA}
\author{W.~B.~Walters}
\affiliation{Department of Chemistry and Biochemistry, University of Maryland, College Park, MD, USA}
\author{A.~W\"{o}hr}
\affiliation{Institute of Structure and Nuclear Astrophysics, University of Notre Dame, Notre Dame, IN, USA}
\affiliation{Joint Institute for Nuclear Astrophysics, University of Notre Dame, Notre Dame, IN, USA}

\date{\today}

\begin{abstract}
\bf{Background:} Nuclear structure plays a significant role on the rapid neutron capture process (r-process) since shapes evolve with the emergence of shells and sub-shells. There was some indication in neighboring nuclei that we might find examples of a new $N$=56 sub-shell, which may give rise to a doubly magic $^{90}_{34}$Se$_{56}$ nucleus.  \bf{Purpose:} $\beta$-decay half lives of nuclei around $^{90}$Se have been measured to determine if this nucleus has in fact a doubly-magic character. \bf{Method:} The fragmentation of a $^{136}$Xe beam at the National Superconducting Cyclotron Laboratory at Michigan State University was used to create a cocktail of nuclei int eh $A$=90 region. \bf{Results:} We have measured the half lives
of twenty-two nuclei near the r-process path in the $A$=90 region. The half lives of $^{88}$As and $^{90}$Se have been measured for the first time. The values were compared with theoretical predictions in the search for nuclear-deformation signatures of a $N$=56 sub-shell, and its possible role in the emergence of a potential doubly-magic $^{90}$Se. The impact of such hypothesis on the synthesis of heavy nuclei, particularly in the production of Sr, Y and Zr elements was investigated with a weak r-process network.
\bf{Conclusions:}  The new half lives agree with results obtained from a standard global QRPA model used in r-process calculations, indicating that $^{90}$Se has a quadrupole shape incompatible with a closed $N$=56 sub-shell in this region. The impact of the measured $^{90}$Se half-life in comparison with a former theoretical predication associated with a spherical half-life on the weak-r-process is shown to be strong.

\end{abstract}

\pacs{23.40.-s; 21.10.Tg; 27.60.+j; 26.30.-k}


\maketitle \section{Introduction}~\label{sec:Introduction}
The rapid neutron-capture process (r-process) is thought to be responsible for the origin of more than half of all nuclei beyond iron. While large uncertainties remain about the site of the r-process and its exact location, it is thought to occur in environments with a high density of free neutrons. There are a number of astrophysical scenarios suggested as possible sites. These include the neutrino driven wind from core-collapse supernovae~\cite{Woo94,Tak94}, two neutron-star mergers~\cite{Fre99}, gamma-ray bursts~\cite{Sur06}, black-hole neutron-star mergers~\cite{Sur08}, relativistic jets associated with failed supernovae~\cite{Sur08,Fuj06} or magnetohydrodynamic jets from supernovae~\cite{Nis06}. 
The speed of the r-process reaction sequence and the resulting abundances of the elements are significantly influenced by the $\beta$-decay properties of the nuclei involved, which are in turn directly related to the evolution of nuclear structure and 
deformation. 
In a recent measurement of the half life of the doubly-magic nucleus $^{78}$Ni~\cite{Hos05,Hos10}, it was shown that the persistence of the closed neutron and proton shells had a significant impact on the r-process abundance patterns, and theoretical predictions for the half life were significantly longer than observed. 

Recent observations of anomalously large abundances of stable Sr, Y and Zr in some metal-poor stars, as compared to heavier neutron-capture elements~\cite{Hon06}, have brought about new questions regarding the r-process mechanism and possible site(s). Travaglio \emph{et al.}~\cite{Tra04} have interpreted these observations as evidence for
a new neutron-capture light-element primary process (LEPP) responsible for the nucleosynthesis of Sr$-$Zr elements. However, the nature of this process remains unclear.
Montes \emph{et al.}~\cite{Mon07} extracted an abundance pattern for the LEPP and demonstrated that it can be associated with different origins. Qian and Wasserburg suggested a charge-particle reaction process (CPR)~\cite{Qui07}, 
which could operate in neutrino-driven winds, 
even in instances when conditions for an r-process are not achieved.
Farouqui \emph{et al.}~\cite{Far09,Kra08} proposed that Sr$-$Zr elements can be primarily produced in the low-entropy expanding envelopes of core-collapse supernovae explosions through charge-particle and neutron-capture reactions.

The main motivation for the present work was to explore the possible existence of a new $N$=56 sub-shell and 
its potential impact on the r-process. \emph{}Whereas the presence of a $N$=56 sub-shell has been 
suggested for Rb$-$Zr elements, its persistence for lighter isotones needs 
further investigation. If the flow of r-process matter emerging from seed nuclei below Kr were trapped in an hypothetical $N$=56 \textquotedblleft ladder$\textquotedblright$, then the post freeze-out $\beta$ decay of nuclei like $^{88}$Ge, $^{89}$As and $^{90}$Se would immediately translate into an enhanced production of stable Sr, Y, and Zr elements---exactly as it occurs at the $N$=82, 126 shells, leading to the well known $A$$\simeq$130, 195 r-process abundance peaks, respectively.

We have measured the half lives of twenty-two nuclei in the $A$=90 region at the National Superconducting Cyclotron Laboratory (NSCL). The half lives of $^{88}$As and $^{90}$Se are reported for the first time, and more precise measurements are presented for the remaining twenty. 
The results are used as a nuclear-structure probe to study the deformation of the $\beta$-decay mother/daughter system, which in turn can provide first hints of the underlying shell structure (see for instance Refs.~\cite{Hos10,Per09,Sar10,Kra84,Meh96,Sor93}).

In the following sections, the experimental setup and analysis techniques employed in the present work are presented.  The results are then discussed in Sec.~\ref{sec:Discussion} in the framework of the quasi-random phase approximation and the impact of the new measurements on the r-process.

\section{Experiment}\label{sec:Experiment}

\subsection{Production and separation of nuclei}\label{sec:Production}
\begin{figure}[h!]
\begin{center}
\includegraphics[width=8cm]{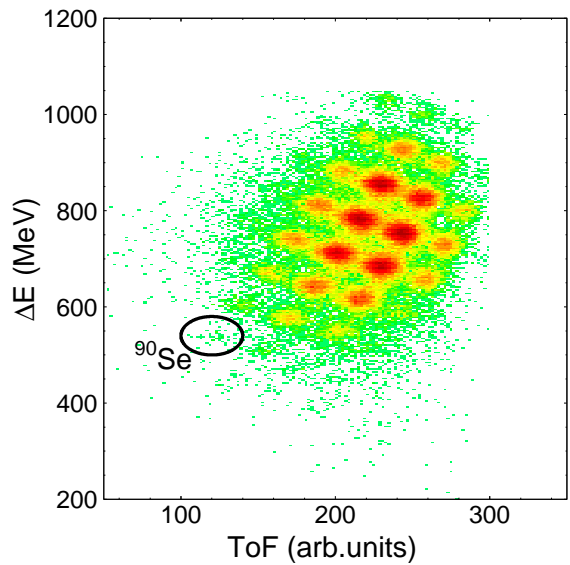} 
\caption{(Color online) Particle-identification (PID) spectum. 
The $ToF$ of the transmitted nuclei is shown versus their energy-loss $\Delta$E in the first Si PIN detector at the experiment station. (See text for details.)}
\label{fig:PID}
\end{center}
\end{figure}

A primary $^{136}$Xe beam was accelerated at the NSCL Coupled Cyclotron Facility~\cite{CCF} to 120 MeV/u, with an average intensity of 1.5~pnA. Nuclei produced by the fragmentation
of the primary beam in  a 235~mg/cm$^{2}$ Be target were forward-emitted into the  A1900 
fragment separator~\cite{Mor03}. The fragments of interest were separated in-flight using the $B\rho$-$\Delta E$-$B\rho$ technique~\cite{Sch87}. The first stage of the separator was set to the magnetic rigidity corresponding to the peak of the momentum distribution of $^{88}$As. An achromatic wedge at the intermediate (dispersive) focal plane was made by combining a BC400 plastic scintillator (I2SCI) (22.22~mg/cm$^{2}$ thickness) with a 22.51~mg/cm$^{2}$ Kapton wedge-foil. The magnets of the second half of the A1900 were set to a value corresponding to the magnetic rigidity of the $^{88}$As fragment after passing through the wedge system. This provided a further filter to select a narrower group of elements. The resulting fragment cocktail was composed of $^{86-88}$As, $^{88-90}$Se, $^{90-93}$Br, $^{91-95}$Kr, $^{93-97}$Rb, $^{95-99}$Sr, $^{97-101}$Y, and $^{100-103}$Zr. The transmitted fragments were uniquely identified event-by-event, by combining their energy-loss measured in the first silicon PIN detector (PIN~1) of the implantation station, and the time-of-flight ($ToF$) measured between I2SCI and a scintillator located in the experimental area, 20~m downstream of the A1900 achromatic focal plane.  To improve the $ToF$ resolution, its dependence on the magnetic rigidity of the transmitted nuclei was corrected using the transversal position in the dispersive image plane measured with I2SCI. Figure~\ref{fig:PID} depicts the particle-identification spectrum (PID) recorded during the course of the experiment for fragments transported the experimental station.

Besides the fragments of interest, the known $\mu$s isomers $^{98}$Y and $^{99}$Y were included in the A1900 setting. These nuclei were used as references to identify all the
fragments transmitted in the same setting. The identification was done at the start of the experiment by implanting nuclei in a 4~mm-thick aluminum foil surrounded by three $\gamma$-ray detectors from the NSCL segmented germanium array (SeGA)~\cite{Mue01}. A PID spectrum shown in Fig. ~\ref{fig:PID} gated on the isomeric $\gamma$ lines 121 keV, 171~keV, and 204~keV from $^{98}$Y provided the reference for the identification of the nuclei in the cocktail. Once the particle identification was established, the implantation foil was removed to transmit particles to the implantation station for the remainder of the experiment.

Since the magnetic rigidity values of the $\mu$s isomers and the fragments of interest were significantly different, the A1900 was set up for the maximum possible $B\rho$ acceptance to allow for their simultaneous transmission. The $B\rho$-window included three primary-beam charge-states $^{136}$Xe$^{+52}$, $^{136}$Xe$^{+51}$ and $^{136}$Xe$^{+50}$, with total intensities exceeding the maxima 1~MHz and 1~KHz tolerated by I2SCI and the Si PIN detectors, respectively. These contaminants were therefore blocked in the first image plane of the A1900 separator with a system consisting of two slits on each side of the beam axis and a central tungsten finger.

\subsection{The $\beta$-decay station}\label{sec:Impsetup}
Nuclei transmitted through the A1900 were implanted in the NSCL Beta Counting System (BCS)~\cite{Pri03}. The BCS was composed by a stack of four silicon PIN detectors (PIN~1--4) of total thickness 2569~$\mu$m, a 979~$\mu$m-thick 40$\times$40 doubly-sided silicon strip detector (DSSD) (40 front-side perpendicular to 40 back-side strips) and a 988~$\mu$m-thick 16-strip single sided Si strip detector (SSSD). A 10~mm-thick Ge crystal downstream of the SSSD was used for light-particle veto purposes. Each PIN detector was connected to a preamplifier with an energy range of 10~GeV. The DSSD strips were connected to dual-gain preamplifiers with ranges equivalent to 3~GeV and 100~MeV, so that signals from implantations and $\beta$ decays could be processed. The 16 SSSD strips were connected to a 16-channel high-gain preamplifier to distinguish $\beta$-decay particles from other particles that produced low-energy signals in the DSSD. The high-gain signals from the DSSD and SSSD preamplifiers were further processed by shaper/discriminator modules designed at Washington University, St. Louis, and manufactured by Pico Systems~\cite{Pico}. These modules include shaper and discriminator circuits sharing the same input, and provide independent logic and analogic output signals. 
The gain of each channel was adjusted in the beginning of the experiment to energy-match the signals from each DSSD strips generated by a $^{228}$Th $\alpha$-source. Independent CFD thresholds were adjusted for each DSSD strip with a $^{90}$Sr $\beta$-decay source. This source was also used to define two thresholds for each SSSD strip, one ($th_{1}$) above the noise peak, and the other ($th_{2}$) just above the $\beta$-decay energy region. Thresholds for each PIN detectors were set on-line using the energy spectra recorded during the experiment.

\begin{figure*}[t!]
\begin{center}
\includegraphics[width=4.2cm]{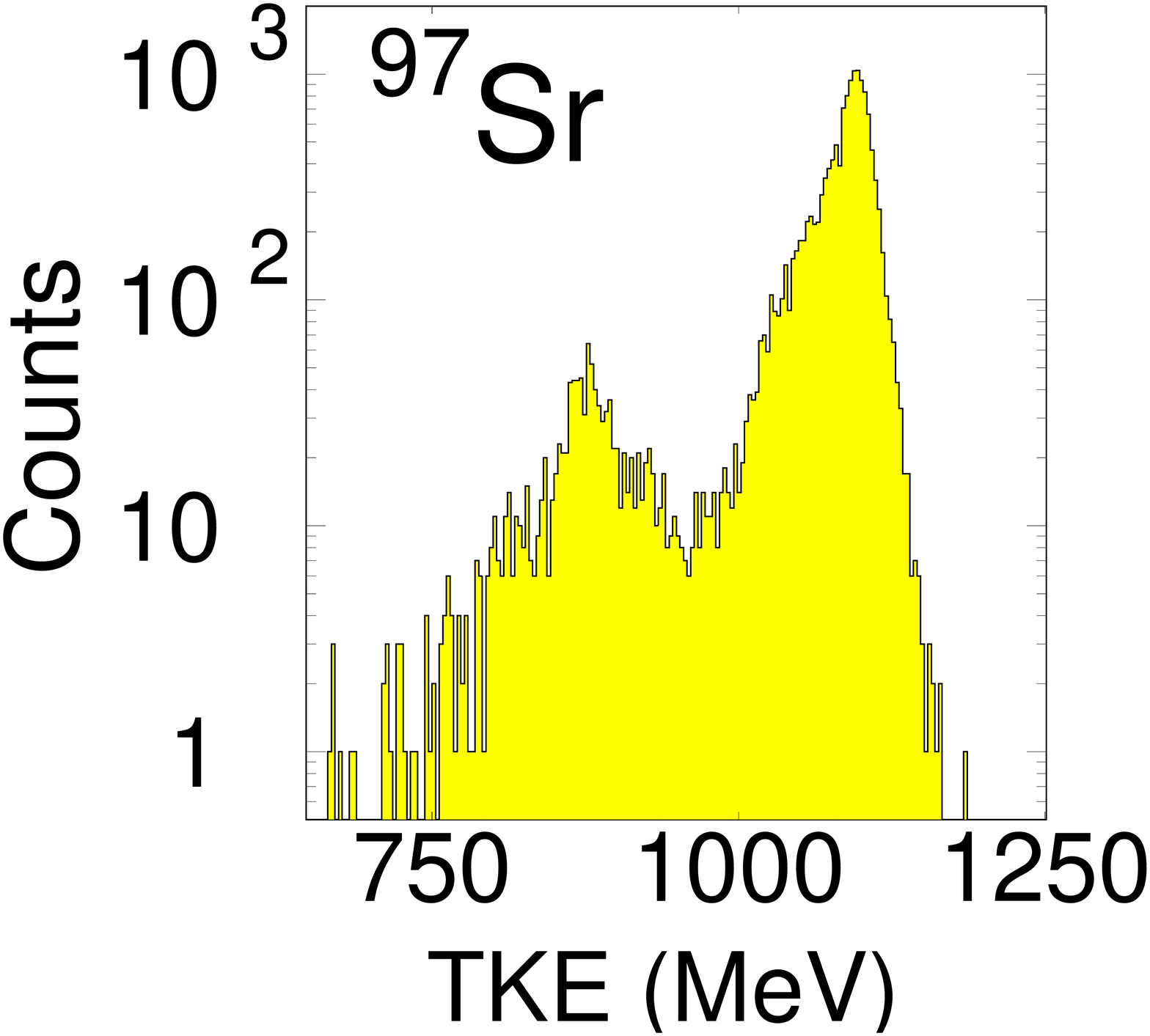}
\includegraphics[width=4.2cm]{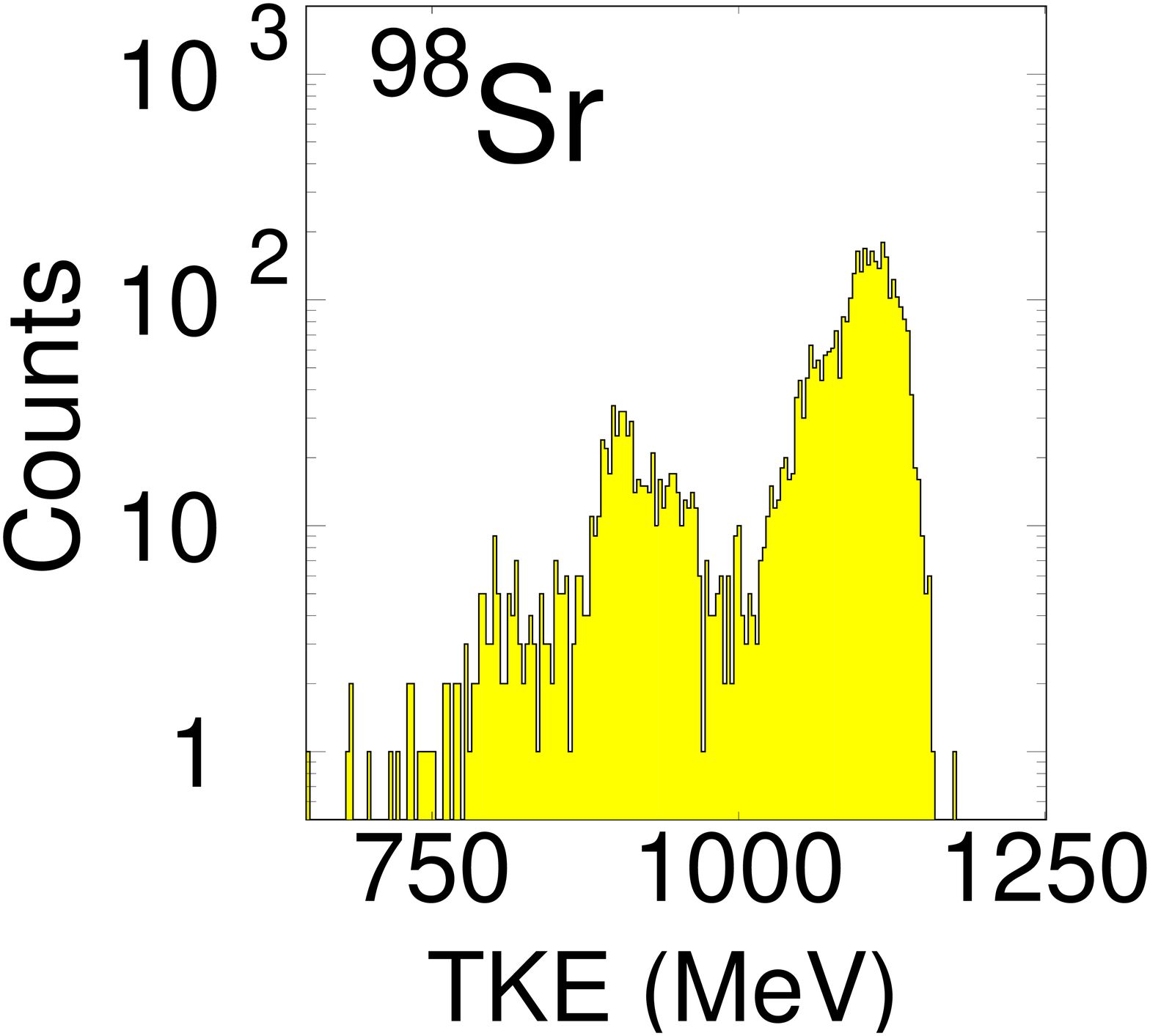}
\includegraphics[width=4.2cm]{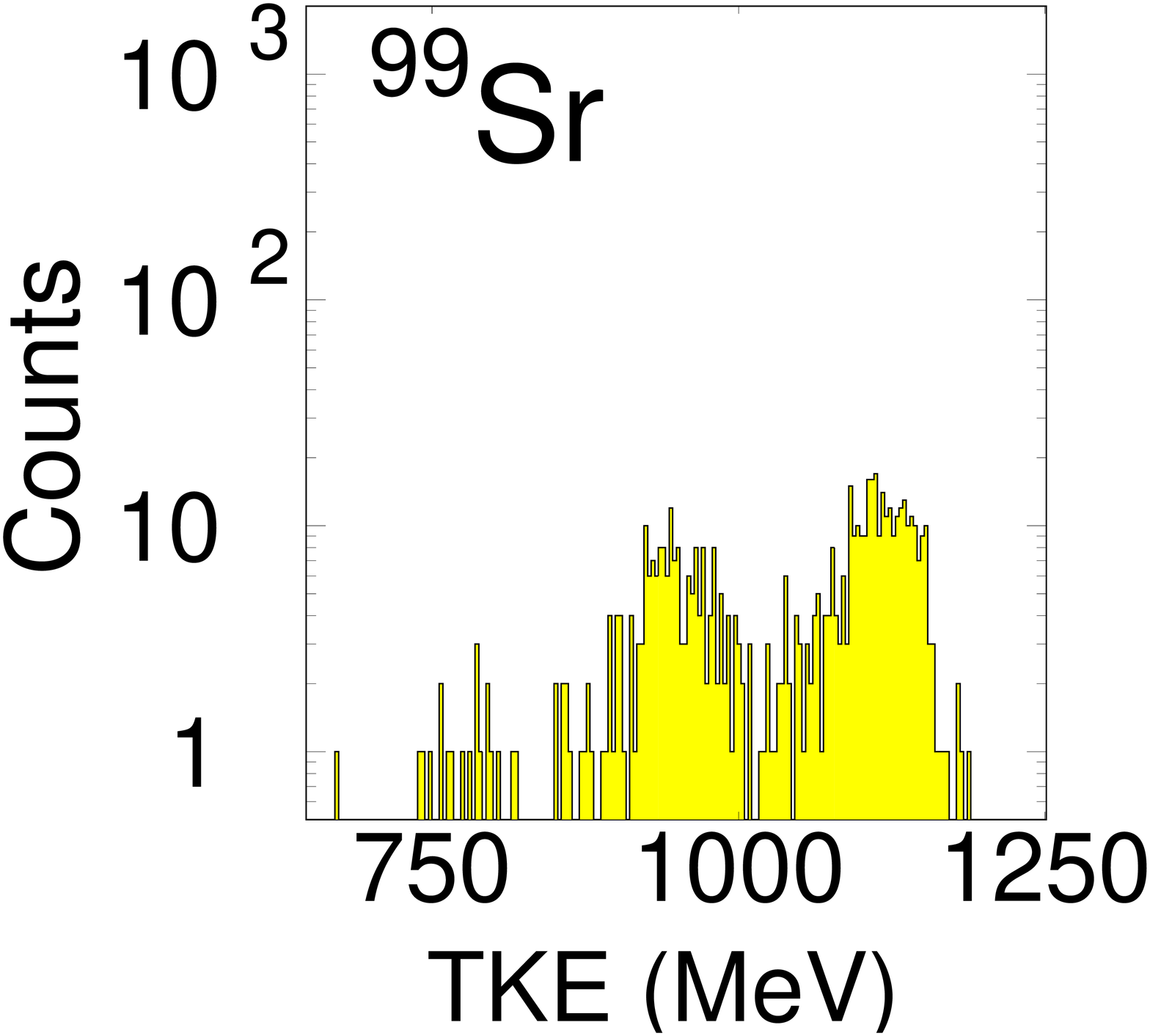} \\
\includegraphics[width=4.2cm]{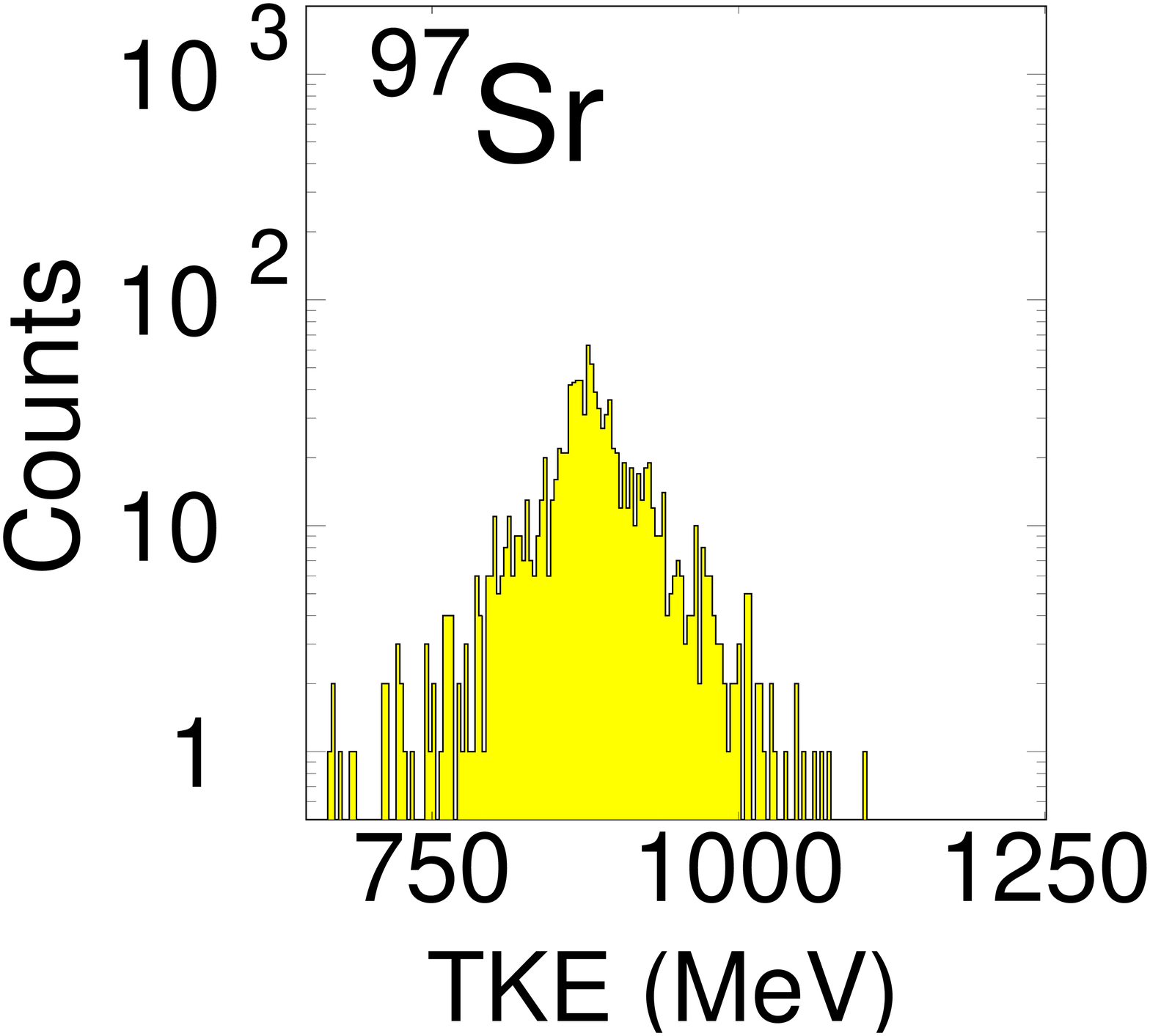}
\includegraphics[width=4.2cm]{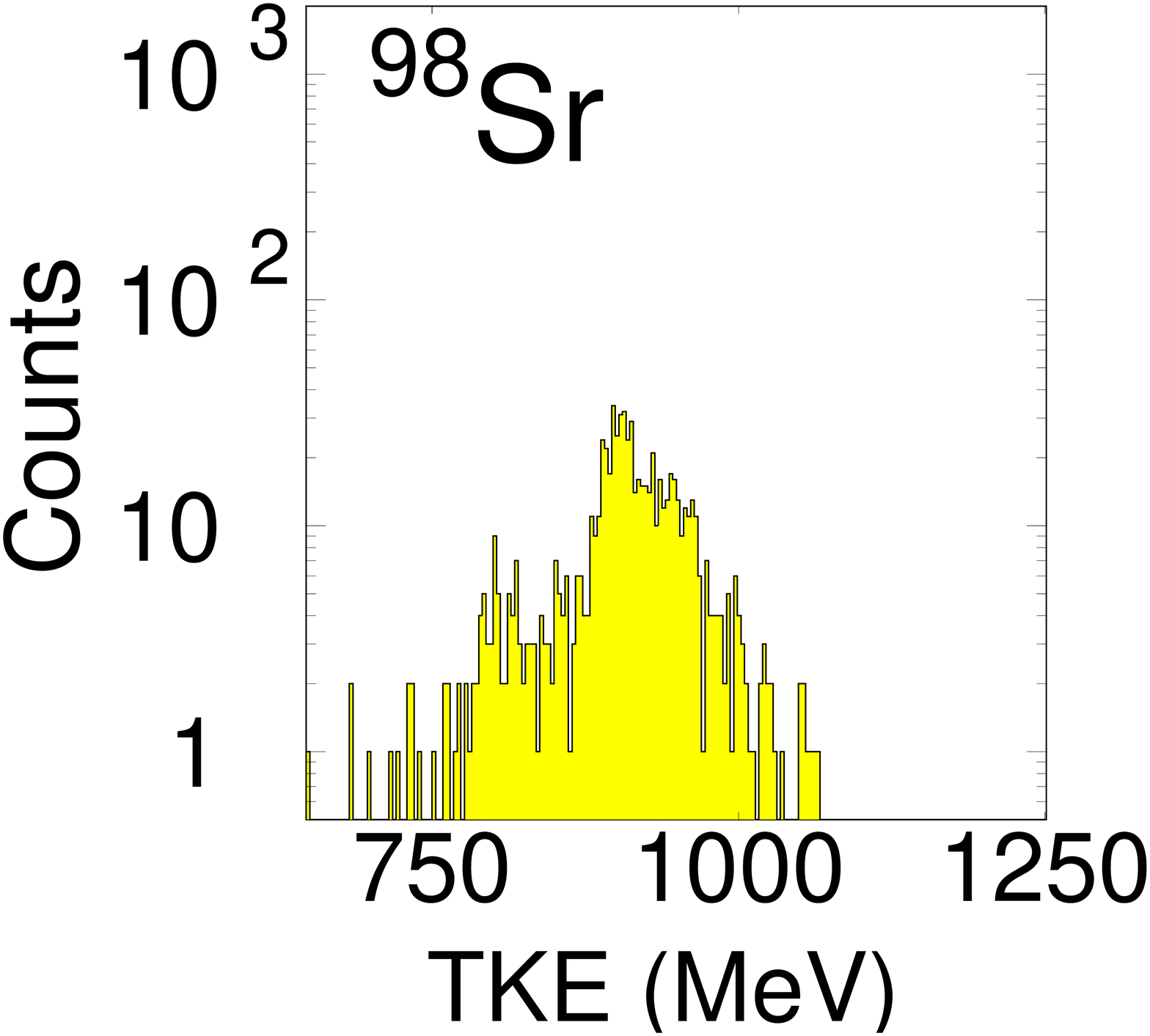}
\includegraphics[width=4.2cm]{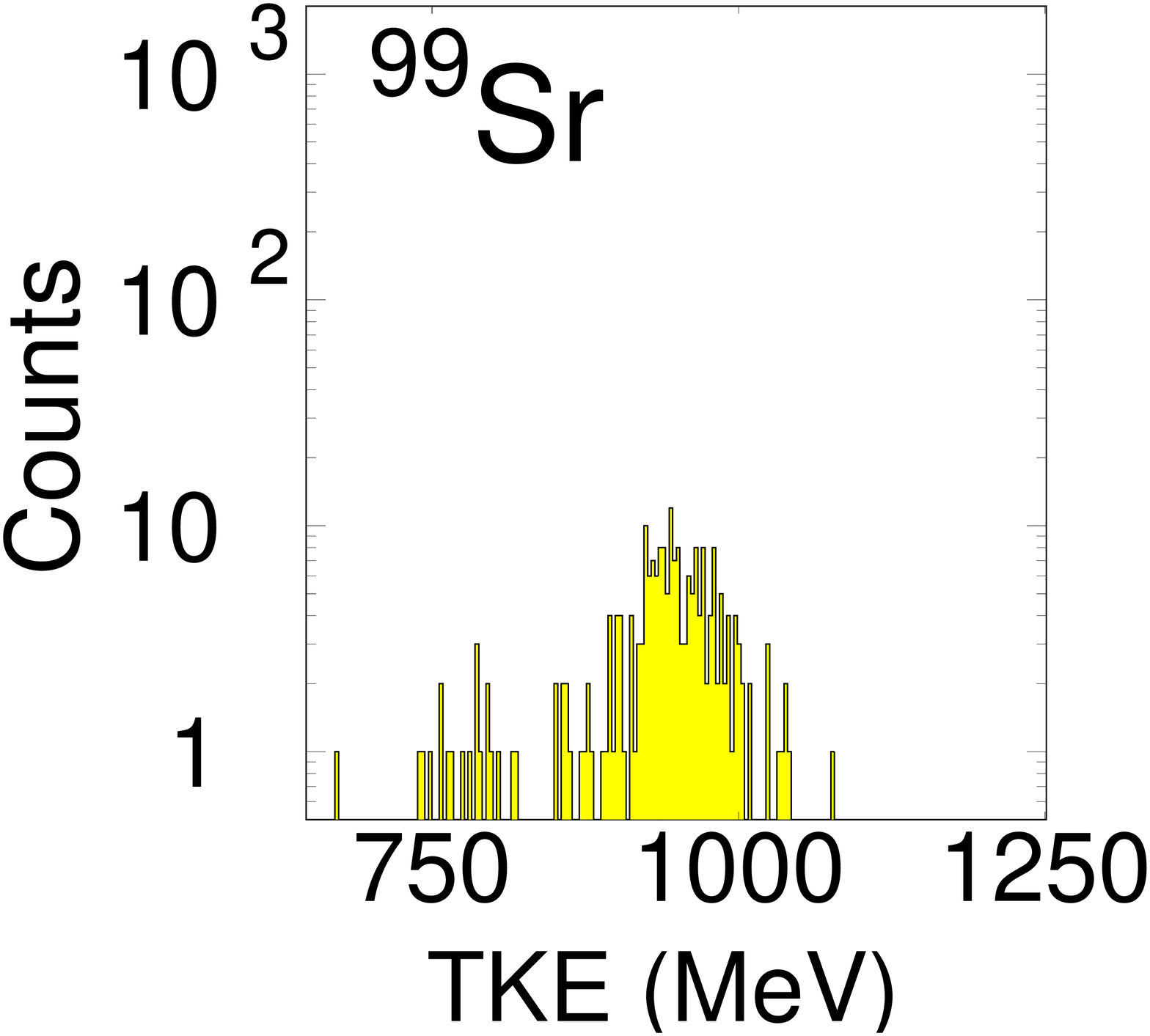} \\
\includegraphics[width=4.2cm]{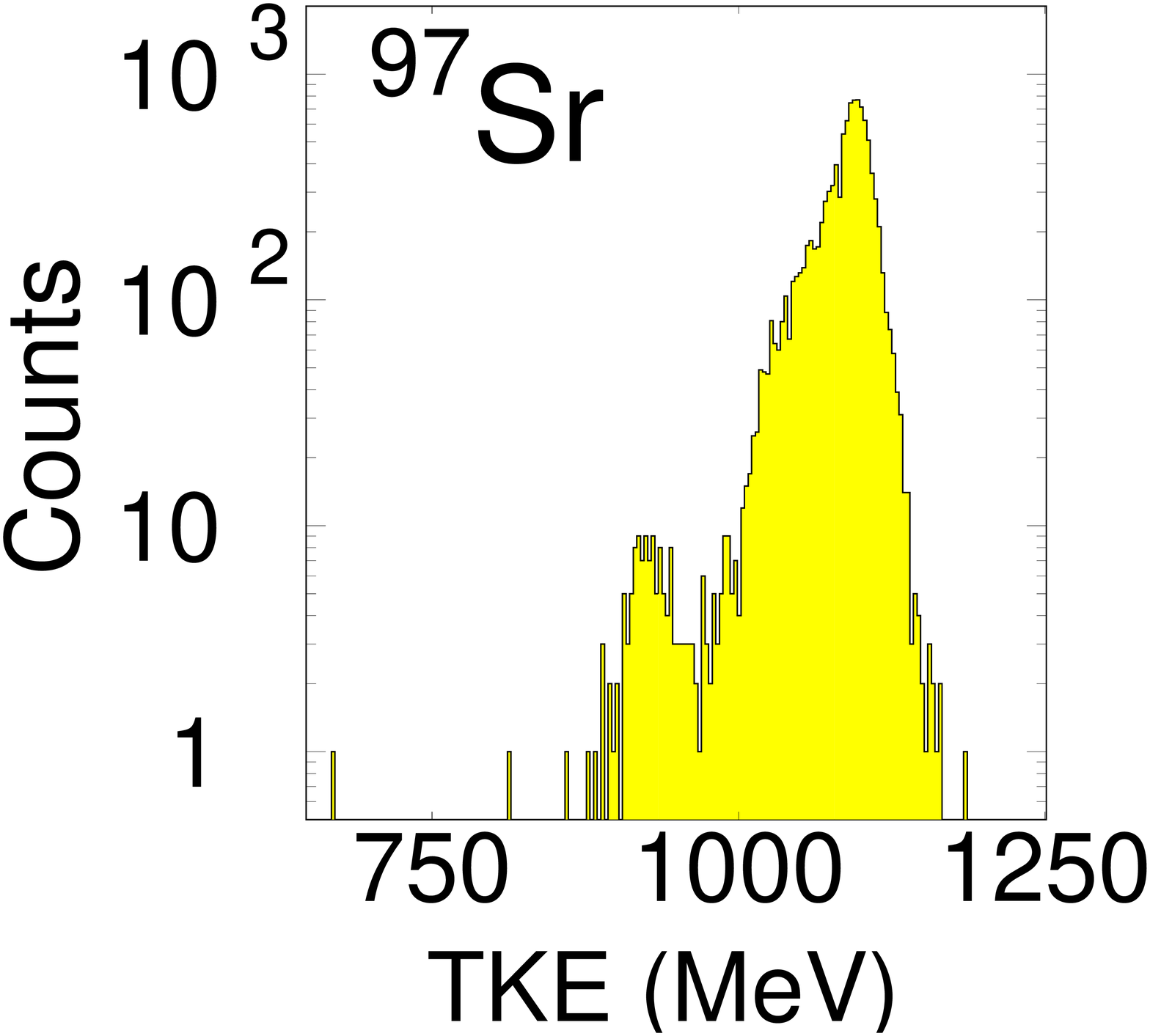}
\includegraphics[width=4.2cm]{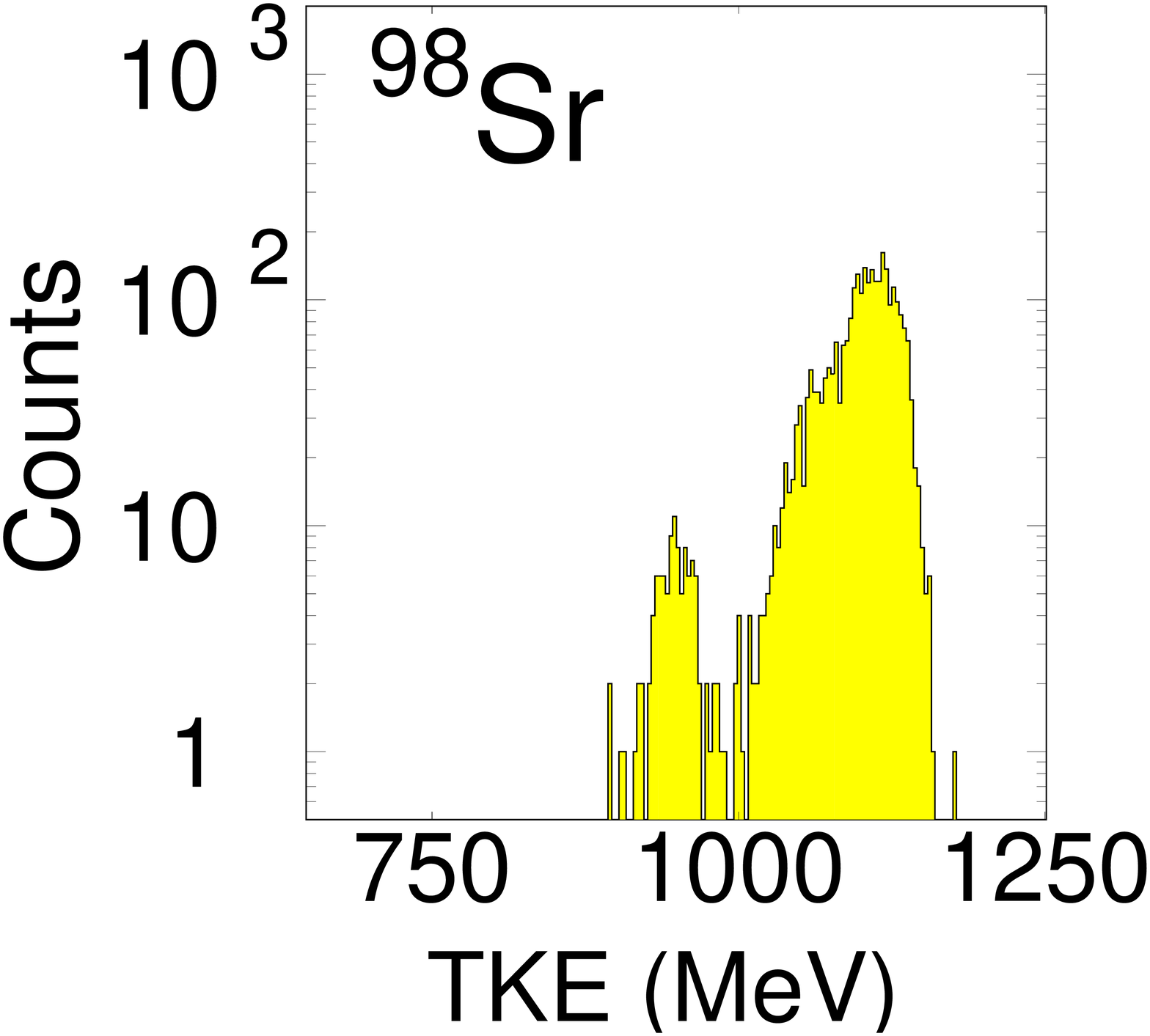}
\includegraphics[width=4.2cm]{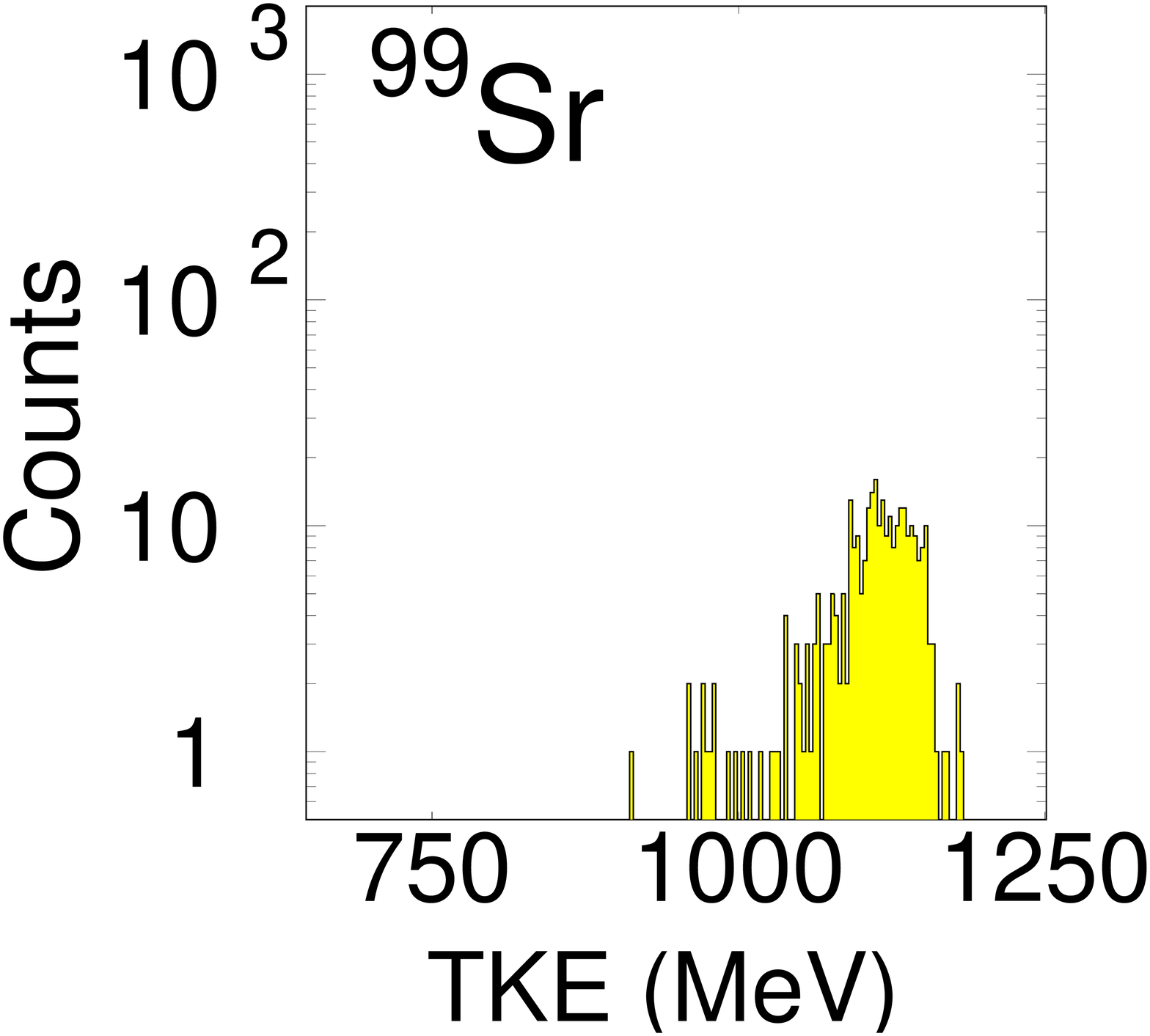} \\
\includegraphics[width=4.2cm]{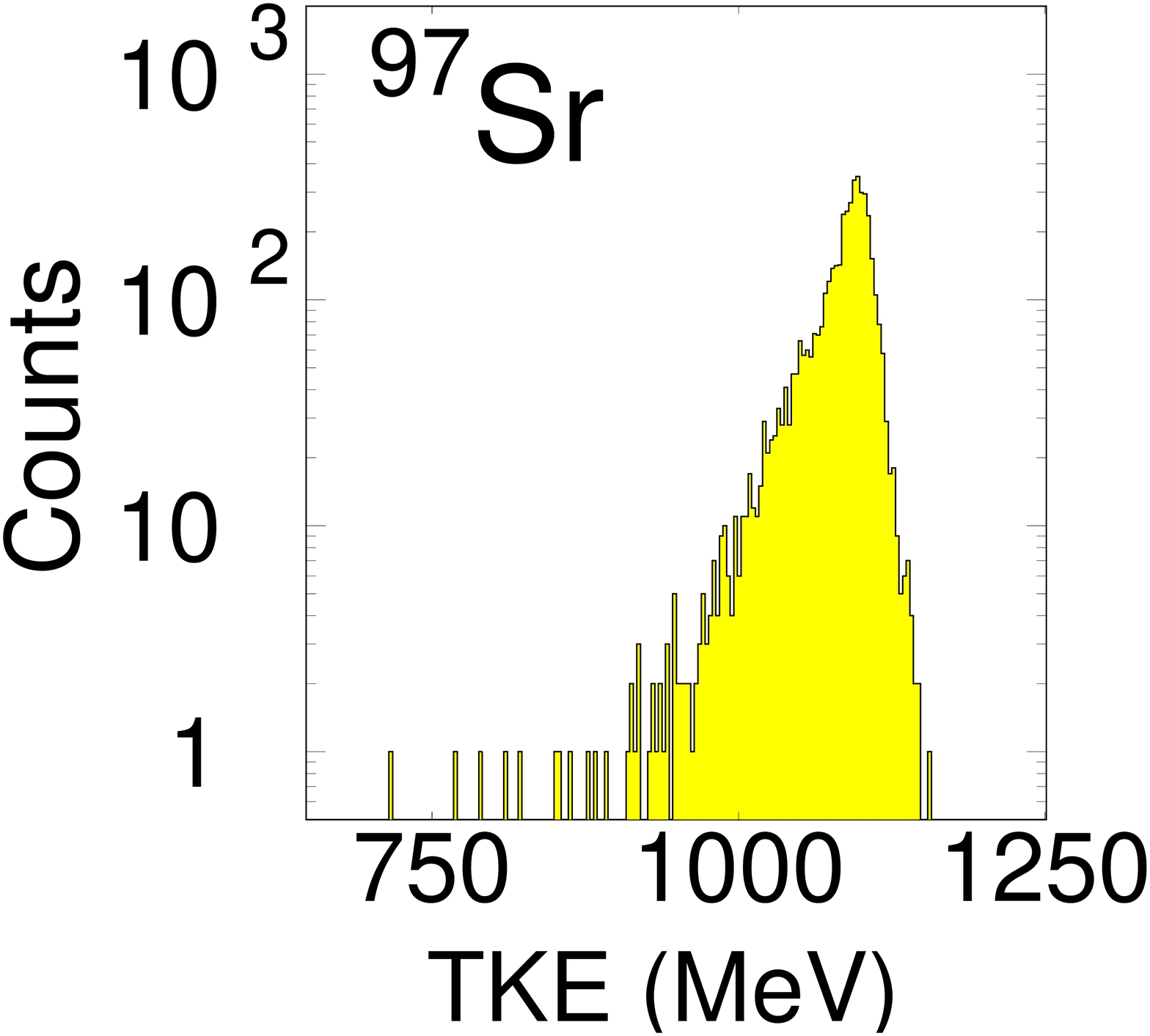}
\includegraphics[width=4.2cm]{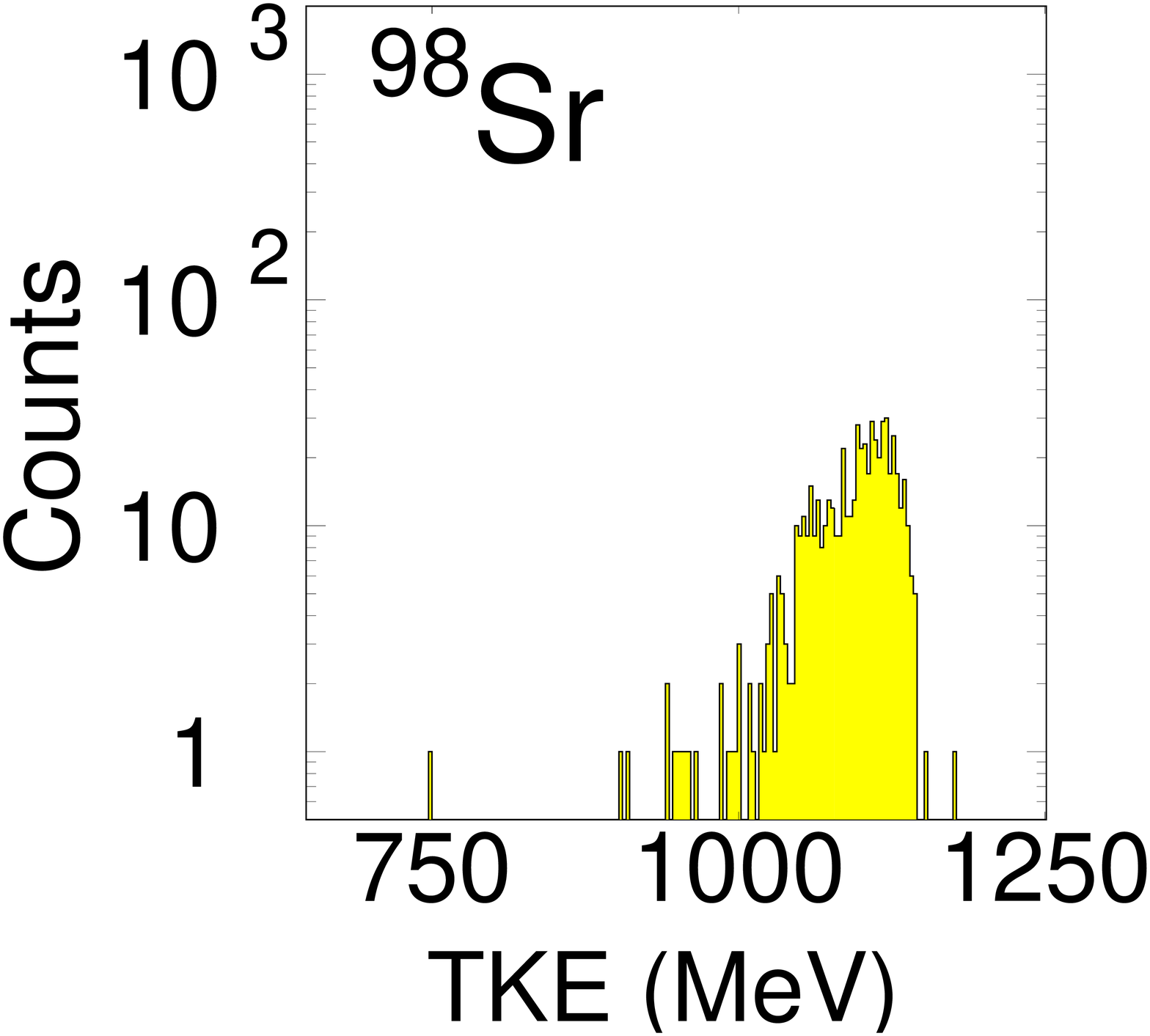}
\includegraphics[width=4.2cm]{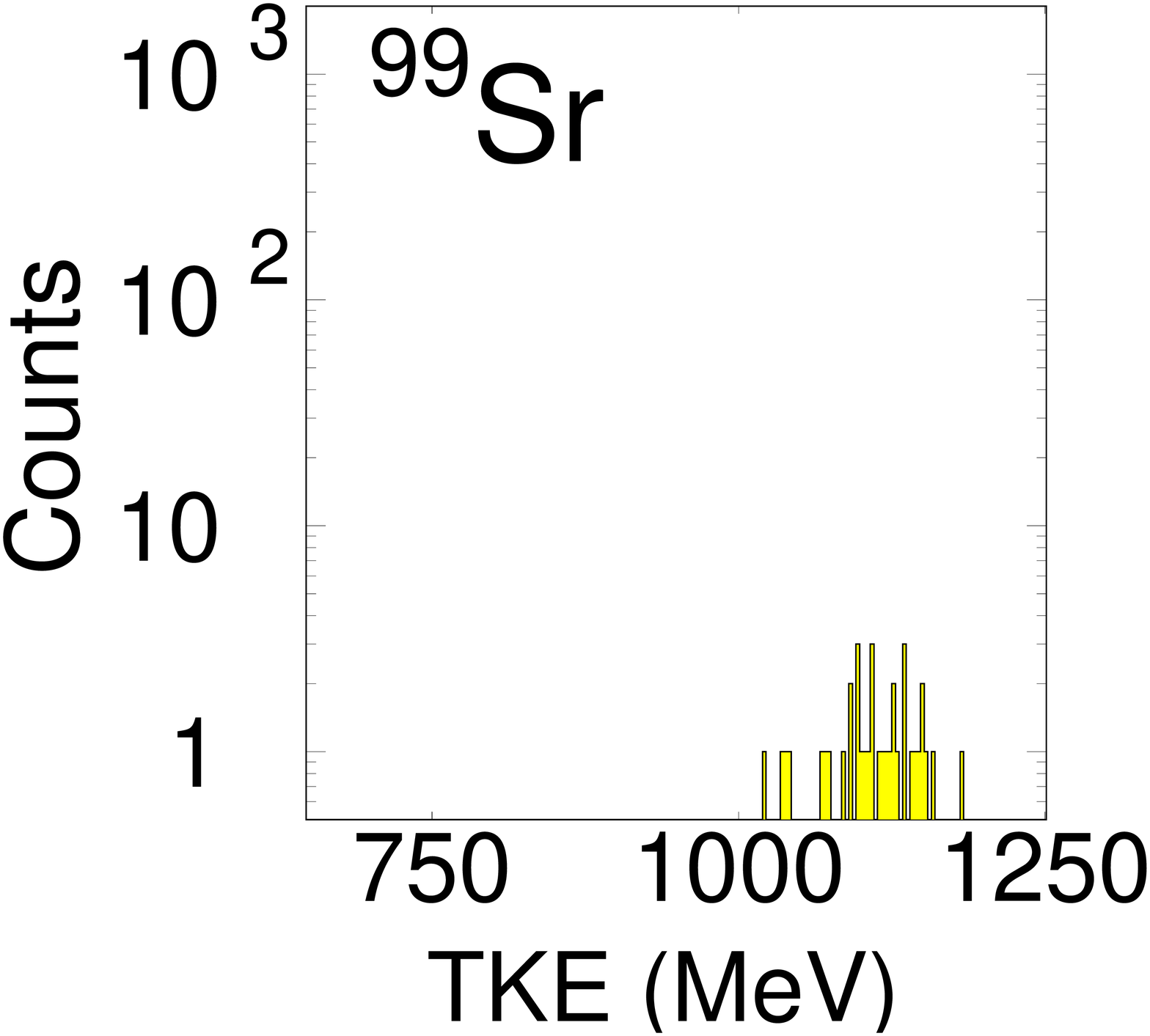} \\
\caption{Total kinetic energy ($TKE$) for  Sr isotopes identified in the PID spectrum. The top row of panels corresponds to all detected events in PIN~1. The second row shows nuclei that were implanted in PIN~4, the third row shows nuclei that were implanted in DSSD and the bottom row shows all nuclei that punched through the DSSD and were consequently lost. The isotope labels correspond to the fully-stripped ions.}
\label{fig:TKE}
\end{center}
\end{figure*}

The DSSD and the four PIN detectors were energy-calibrated by comparing the signals from different transmitted nuclei with the values calculated with the LISE program~\cite{Baz02}, using the Ziegler energy-loss formulation~\cite{Zie85}. A dedicated un-wedged setting of the A1900, transmitting a large number of different fragments, was used in the beginning of the experiment to cover as broad an energy range as possible. The total kinetic energy ($TKE$) of each transmitted ion was determined by summing the calibrated energy signals from the four PIN detectors and the DSSD. The $TKE$ was used to determine the charge states of the transmitted ions. Thus, fully-stripped nuclei could be separated from charge-state contaminants---mostly hydrogen-like ions with mass numbers $A-2$ and $A-3$, and $A-5$ helium-like ions---with similar $\Delta E$ and $ToF$ values. An example of the charge-state separation based on $TKE$ is shown in Fig.~\ref{fig:TKE} for Sr isotopes. 
For each isotope, the most intense high-$TKE$ peak arises from the fully-stripped species, whereas the medium- and lowest-$TKE$ peaks correspond to the hydrogen-like and helium-like contaminants, respectively. 
As can be seen, nearly all the charge-state contaminants were mainly implanted in the last PIN. On the other hand, there was a fraction of fully-stripped nuclei of interest that passed through the DSSD and were thus implanted in the SSSD. 
The fraction of these lost events was negligible for the lightest, neutron-rich nuclei of interest.

\section{Data analysis}\label{sec:Analysis}

\subsection{Identification of $\beta$ decays}\label{sec:indentbdec}
An implantation event was defined as a signal in the first PIN detector in coincidence with low-gain output signals from at least one strip on each side of the DSSD, and in anti-coincidence with signals from SSSD above threshold $th_{2}$ (see Section~\ref{sec:Impsetup}). Decay-like events were defined by a high-gain output signal from at least one strip on each side of the DSSD, in anti-coincidence with signals from any PIN detector, SSSD below $th_{1}$ (i.e., in the noise-peak region), and Ge overflows generated by light particles. For each implantation event, the strip location on each side of the DSSD, defining the implanted pixel, was calculated as the average between all the strips weighted by their respective energy-signal amplitude. The implanted pixel and time, registered by a continuously counting 50~MHz clock, were recorded for all implantation events. Subsequent decay-like events occurring in the same or nearest neighbor pixels (defining a cluster of nine pixels) within a given correlation-time window $t_{c}$ were associated with the previous implantation, and their times and pixels recorded. If a decay-like event was time correlated with more than one implantation measured in the same 9-pixel cluster (back-to-back implants), then all the events within the sequence were rejected. Data were taken in 1-hour runs over a total running time of 140 hours.

Several sources contributed to the $\beta$-decay background. The most important were light particles that did not lead to overflows in the Ge detector and  had energies in SSSD below $th_{2}$ and above $th_{1}$ (i.e., in the $\beta$-like event region); real $\beta$ decays from longer-lived nuclei implanted in the same pixel; and electronic noise signals above the thresholds. The number of $\beta$-like events in each 9-pixel cluster occurring at a time $t_{b}$ after a preceding implantation was recorded for each 1-hour run time to quantify the $\beta$-decay background. 
The optimum value of $t_{b}$ was determined from an analysis of the total background rate $R_{b}$ as a function of $t_{b}$. If $t_{b}$ was too short, $R_{b}$ included background events and real correlated $\beta$ decays. As $t_{b}$ increased, the number of real correlated $\beta$ decays contributing to $R_{b}$ was lower, and $R_{b}$ decreased with $t_{b}$. Beyond a given value of $t_{b}$, no correlated $\beta$ decay were included and $R_{b}$ remained constant. The optimum value was found at $t_{b}$=40~s.
The background rate of each 9-pixel cluster and run was used to compute the total background rate of each isotope. The statistical error derived from the number of background events recorded in each DSSD cluster and run was about 8$\%$. An average background rate, among the different nuclei analyzed, of about 0.01~s$^{-2}$ was found with a variance within the statistical error. The number of background events $B_{\beta}$ associated with a given isotope was calculated by multiplying its background rate by the number of implants and the correlation time $t_{c}$.

\subsection{$\beta$-decay half-lives}\label{sec:halflive}
Up to three correlated $\beta$-decay events were accepted for each implantation, defining a $\beta$-decay chain. Possible members of these chains included the $\beta$ decay of the daughter and granddaughter; the $\beta$ decay of the daughter and granddaughter generated by $\beta$-delayed neutron emission from the implanted nucleus; and the $\beta$ decay of the granddaughter following the $\beta$-delayed two-neutron emission from the $\beta$-decay daughter. For each implantation, the decay times $t_{i}$ (with $i$=1, 2 or 3) of the three $\beta$-decay chain members were recorded along with the background rate in the corresponding cluster. All the decay chains of a given nucleus were analyzed using the Maximum Likelihood Method (MLH) described in Ref.~\cite{Per09}.


The $\beta$-decay detection efficiency $\epsilon_{\beta}$ was deduced for each nucleus as the ratio of the number of detected $\beta$ decays ($N_{\beta}$) coming from the mother implanted nucleus to the number of implantations. The resulting value was nearly constant among all the nuclei considered. A weighted average efficiency (29$\pm$5)$\%$ was used in the MLH fits. $N_{\beta}$ was
given by $N_{\beta}=N_{0}/(\lambda_{1} \Delta t)$, where $\lambda_{1}$ (the decay constant of the mother nucleus) and $N_{0}$ (the initial number of mother decaying nuclei) were the fit free parameters, and $\Delta t$ was the bin size of the decay-curve histogram. As discussed in Ref.~\cite{Per09}, the total number of events per bin $\Delta t$ in the decay curve must be sufficiently high (i.e. above $\simeq$10) for the above method to be statistically justified. The fit function was derived from the Batemann equations~\cite{Cet06}, including constant background rates, and the $\beta$ decay of the mother and all descendants.



The most important sources of error in the determination of the half lives were the uncertainty in the background rate, $\epsilon_{\beta}$, and the decay parameters of the descendant nuclei. The systematic error was determined from the different MLH fits varying these parameters randomly within their uncertainty range. The average systematic errors from background, efficiency and decay parameters were 2.9\%, 5.2\% and 0.5\%, respectively, yielding a total average systematic error of about 6.6\%. The statistical error was calculated according to W.~Br\"uchle~\cite{Bru03}.

\begin{table*}[t]
\caption{Total number of implantations, number of decay events per histogram $N$ and experimental $\beta$-decay half lives obtained from the Maximum Likelihood method (MLH) and least-squares fits (Least squares), including systematic and statistical errors. The results are compared with known half-life values~\cite{ENSDF}.}
\begin{ruledtabular}
\begin{tabular}{cccccc}
&  &  &  &  & \\
Isotope    &    Implantations   &  $N$ & \multicolumn{3}{c}{Half-life (ms)} \\
&  &  &  &  & \\
    \cline{4-6}
&  &  &  This work& This work & \\
&  &  & MLH & Least squares & Literature
\\ \hline
&  &  &  &  & \\\vspace{1mm}
$^{87}$As  &  27   &  12  & $1450(550)^{+3900}_{-1250}$  &                & 560(80)       \\\vspace{1mm}
$^{88}$As  &  16   &  8   & $200(10)^{+200}_{-90}$       &                &               \\\vspace{1mm}
$^{88}$Se  & 144   &  74  & $650(35)^{+175}_{-140}$      &                & 1530(60)      \\\vspace{1mm}
$^{89}$Se  & 180   &  90  & $345(25)^{+95}_{-80}$        &                & 410(40)       \\\vspace{1mm}
$^{90}$Se  & 70    &  30  & $195(10)^{+95}_{-65}$        &                &               \\\vspace{1mm}
$^{90}$Br  & 869   & 369  & $1850(130)^{+230}_{-210}$    & 2085(25)(235)  & 1910(10)      \\\vspace{1mm}
$^{91}$Br  & 1158  & 531  & $615(35)^{+65}_{-60}$        & 495(30)(40)    & 541(5)        \\\vspace{1mm}
$^{92}$Br  & 309   & 161  & $290(15)^{+70}_{-55}$        &                & 343(15)       \\\vspace{1mm}
$^{93}$Br  & 45    & 20   & $69(3)^{+41}_{-25}$          &                & 102(10)       \\\vspace{1mm}
$^{93}$Kr  & 1661  & 831  & $1040(70)^{+110}_{-165}$     & 1245(30)(75)   & 1290(10)      \\\vspace{1mm}
$^{94}$Kr  & 881   & 480  & $282(9)^{+29}_{-27}$         & 275(1)(27)     & 212(5)        \\\vspace{1mm}
$^{95}$Kr  & 152   & 68   & $250(10)^{+75}_{-55}$        &                & 114(3)        \\\vspace{1mm}
$^{95}$Rb  & 7587  & 3253 & $410(20)(25)$                & 412(1)(13)     & 378(1)        \\\vspace{1mm}
$^{96}$Rb  & 1972  & 1904 & $220(9)(17)$                 & 212(6)(16)     & 203(3)        \\\vspace{1mm}
$^{97}$Rb  & 291   & 164  & $208(6)^{+42}_{-36}$         &                & 169(1)        \\\vspace{1mm}
$^{96}$Sr  & 10520 & 5385 & $940(50)(60)$                &  950(25)(35)   & 1070(10)      \\\vspace{1mm}
$^{97}$Sr  & 9763  & 5082 & $450(25)(30)$                &  456(5)(13)    & 429(5)        \\\vspace{1mm}
$^{98}$Sr  & 2545  & 1377 & $640(25)(35)$                &  577(2)(30)    & 653(2)        \\\vspace{1mm}
$^{99}$Sr  & 231   & 131  & $420(20)^{+115}_{-90}$       &                & 269(1)        \\\vspace{1mm}
$^{99}$Y   & 6061  & 3394 & $1355(70)(85)$               &  1400(15)(70)  & 1470(7)       \\\vspace{1mm}
$^{100}$Y  & 1699  & 878  & $840(60)(80)$                &  845(45)(75)   & 735(7)        \\\vspace{1mm}
$^{101}$Y  & 239   & 117  & $480(30)^{+140}_{-110}$      &                & 450(20)       \\\vspace{1mm}
\end{tabular}
\end{ruledtabular}
\label{tab:halflives-results}
\end{table*}

\section{Results and discussion}~\label{sec:Discussion}
\begin{figure*}[t!]
\begin{center}
\includegraphics[width=4.2cm]{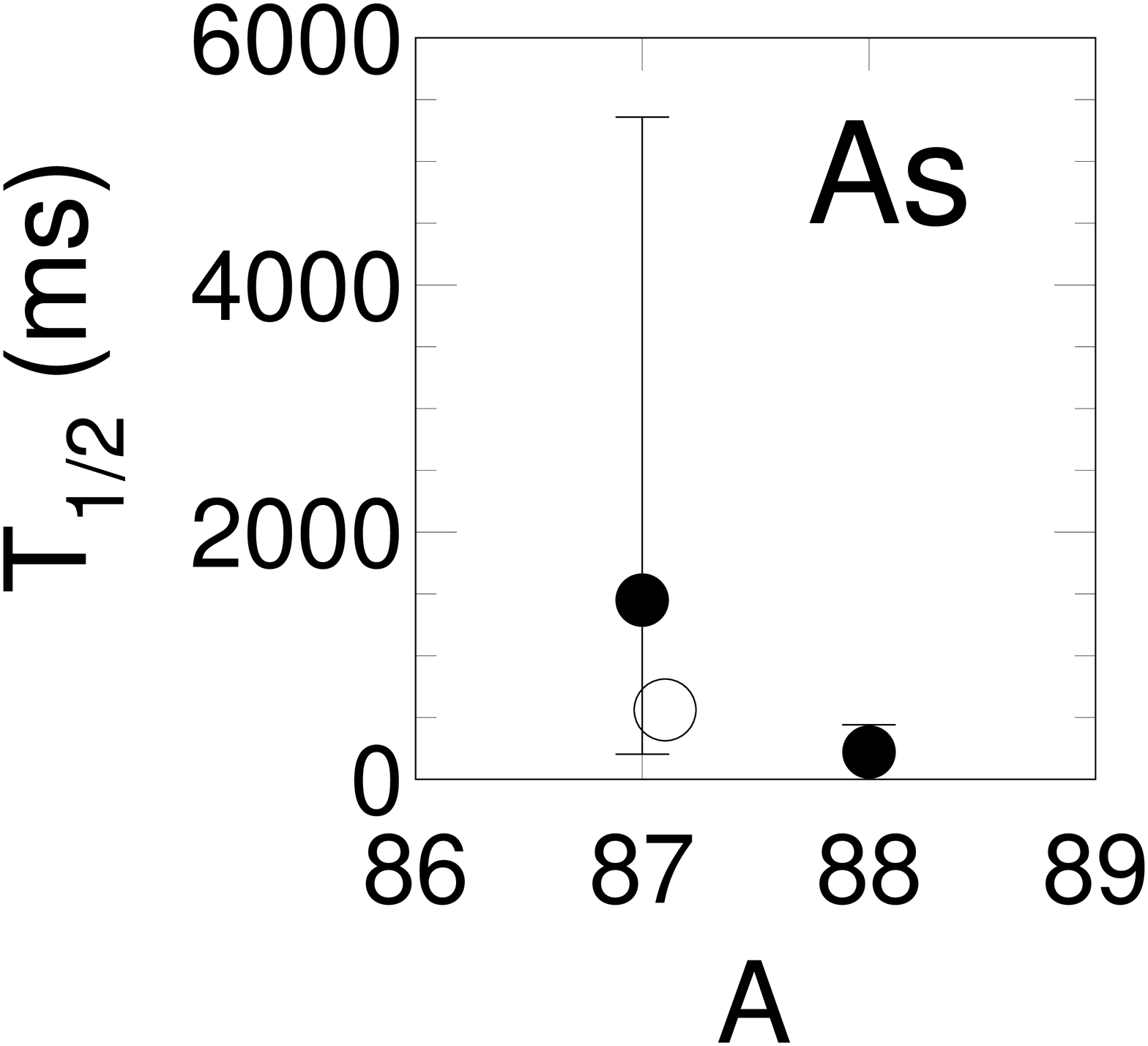}
\includegraphics[width=4.2cm]{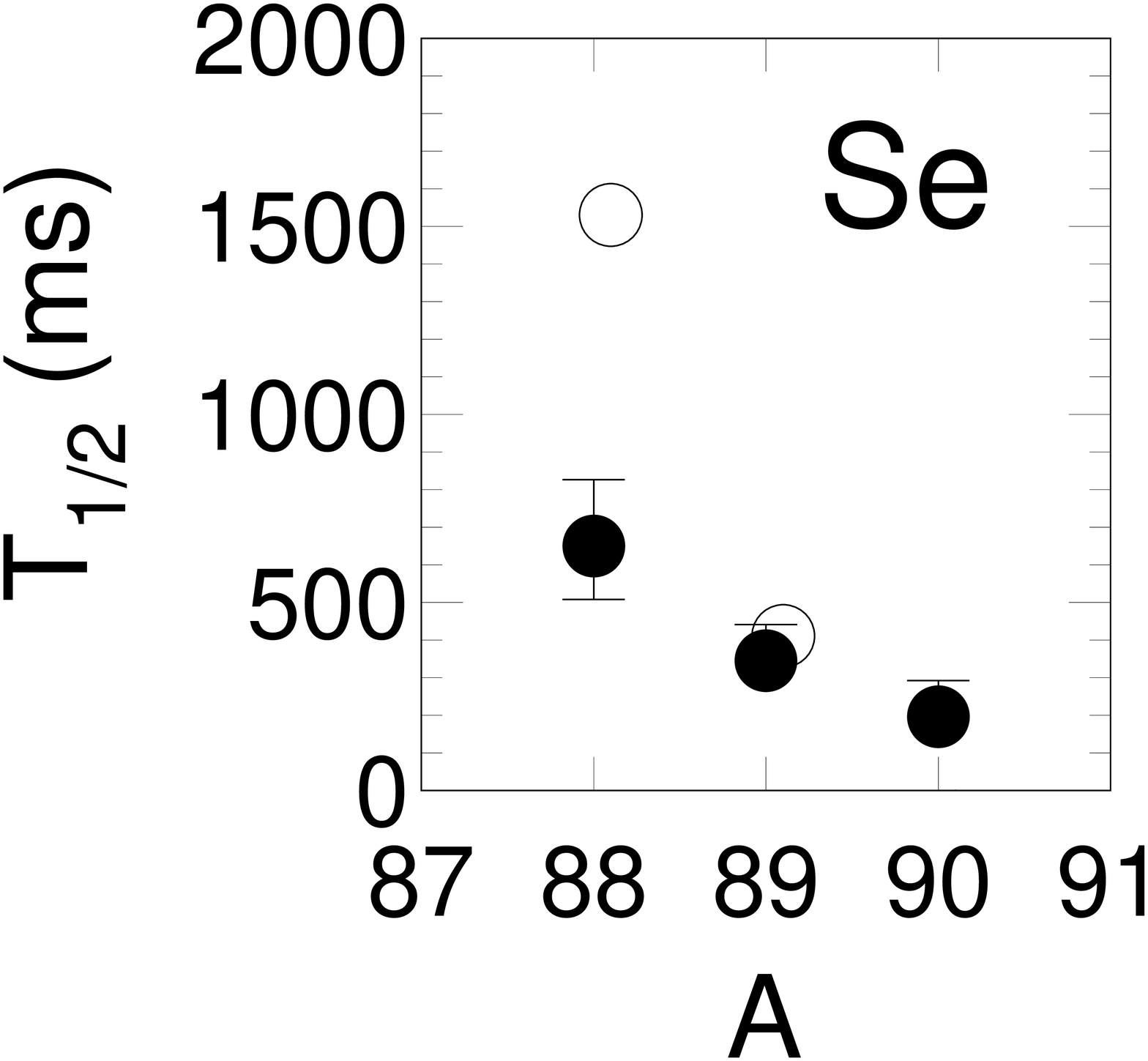}
\includegraphics[width=4.2cm]{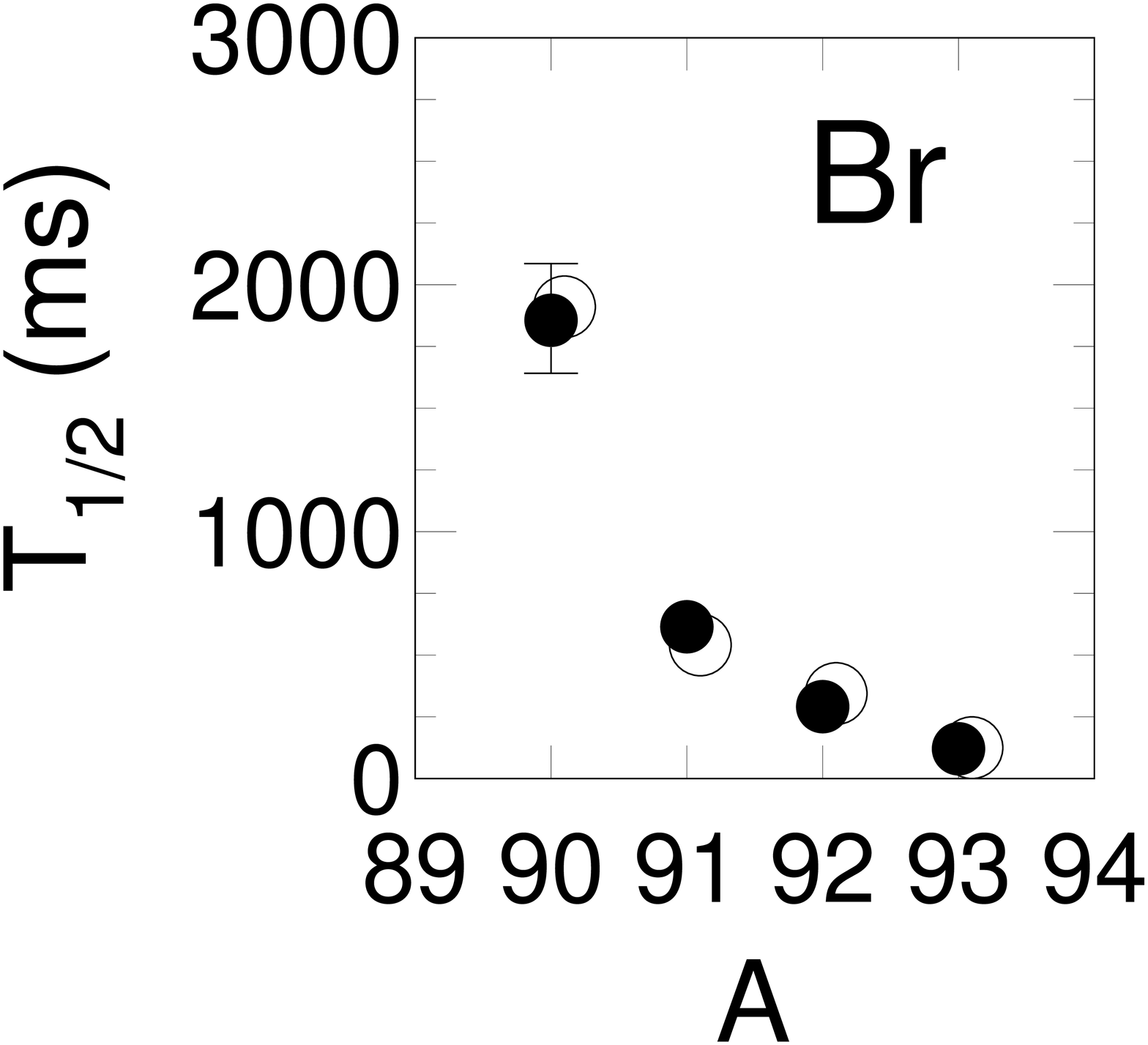}
\includegraphics[width=4.2cm]{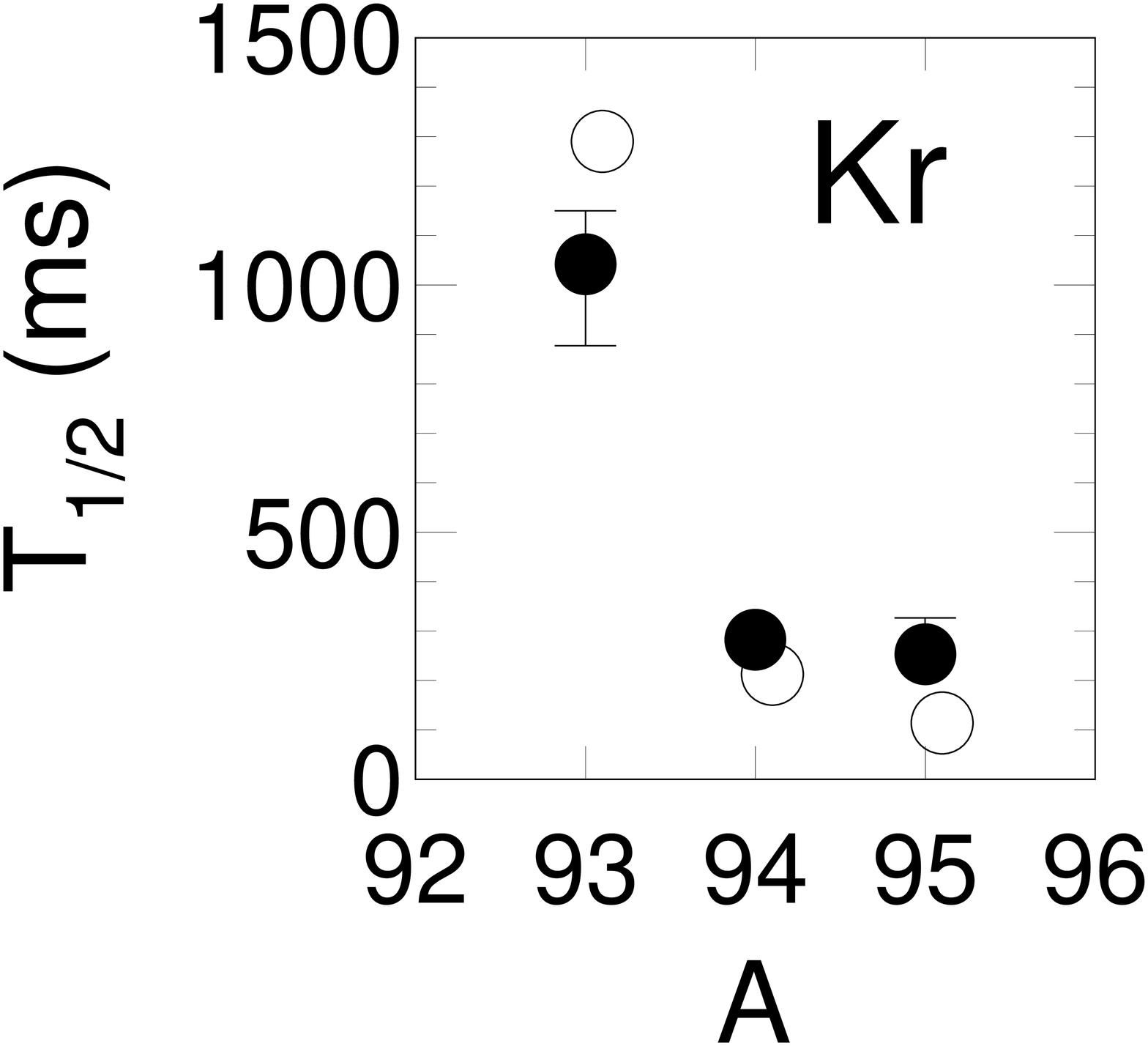} \\
\includegraphics[width=4.2cm]{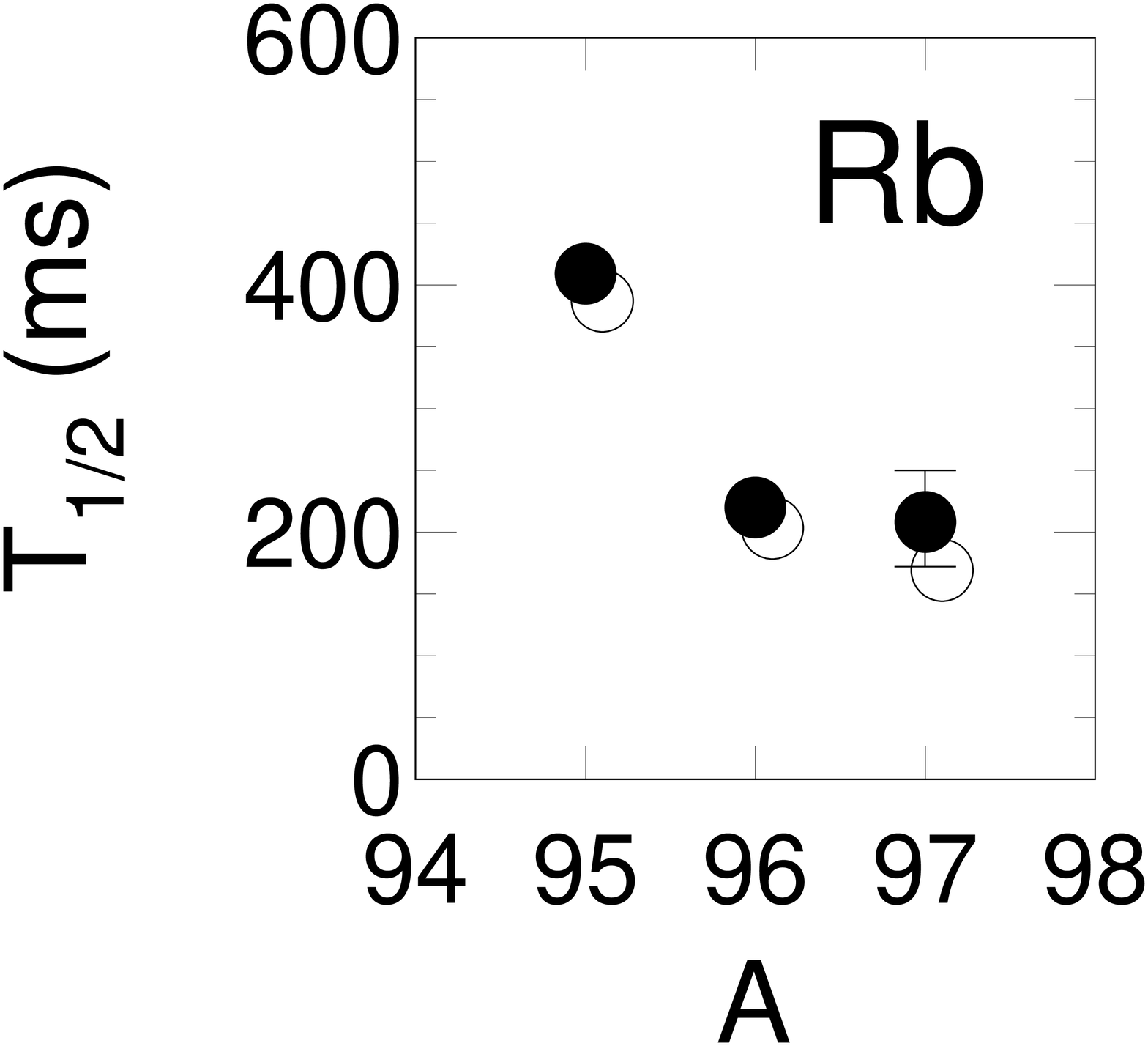}
\includegraphics[width=4.2cm]{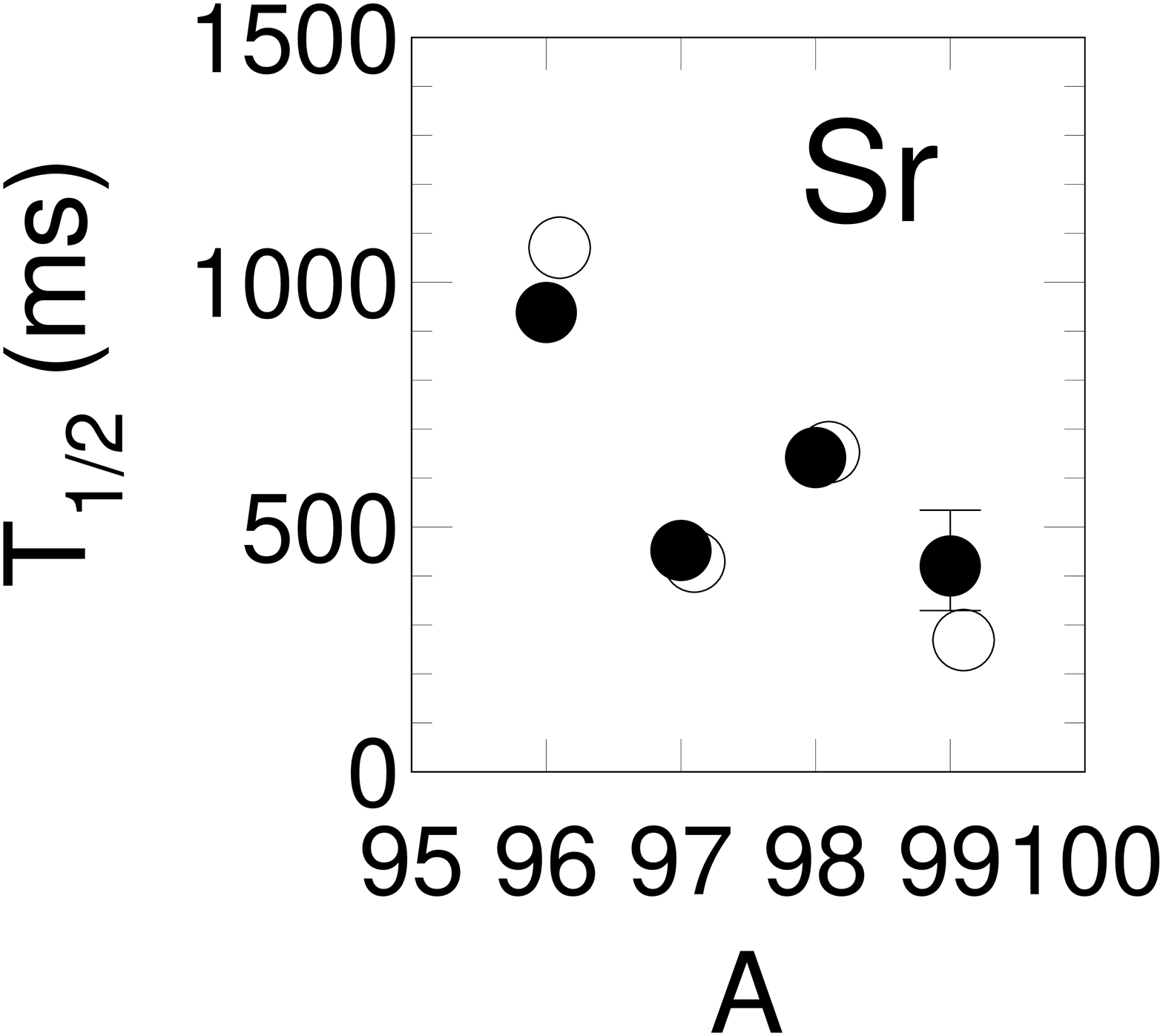}
\includegraphics[width=4.2cm]{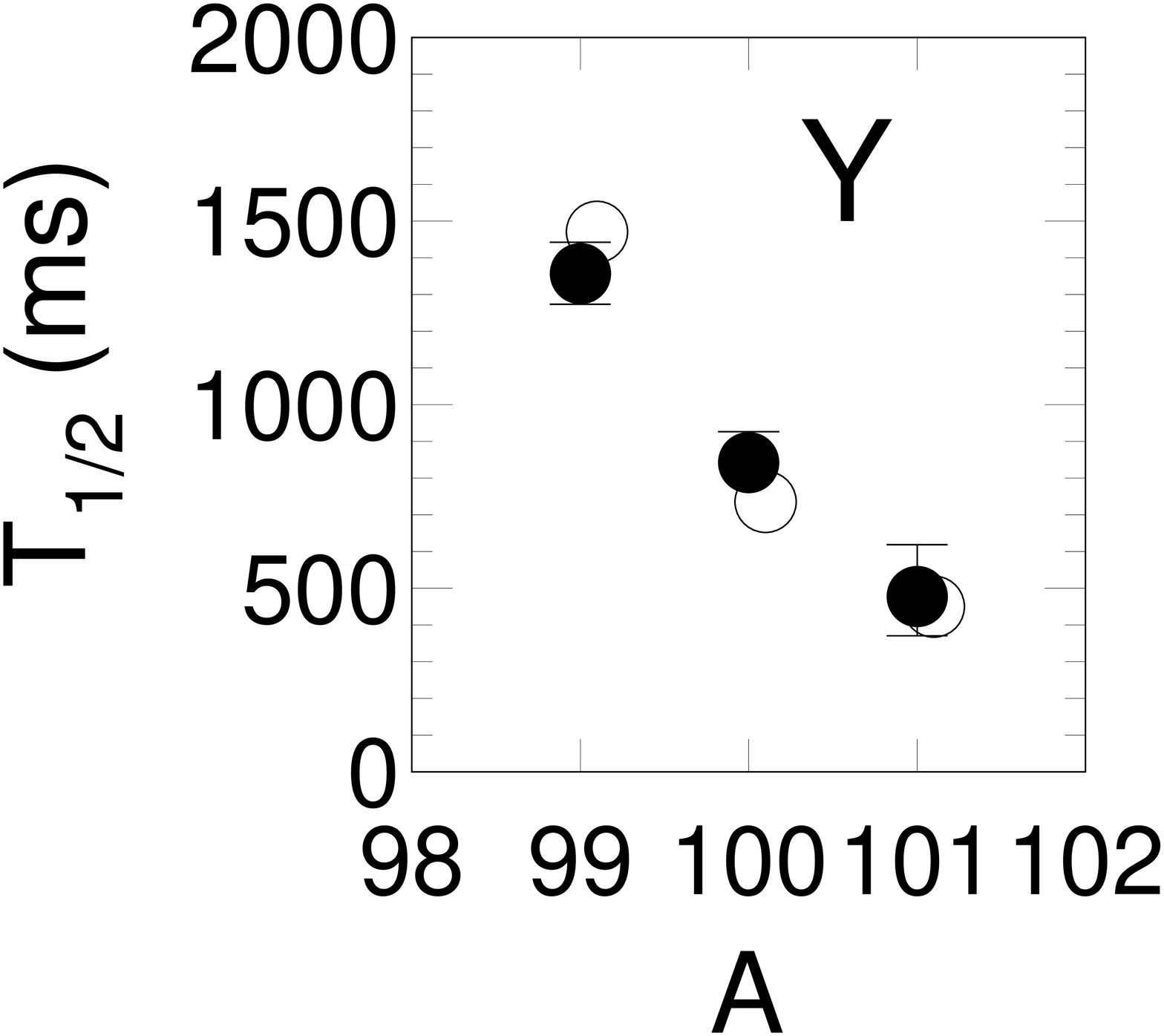}
\caption{$\beta$-decay half-lives obtained in this work from the Maximum Likelihood method for the As, Se, Br, Kr, Rb, Sr, and Y isotopes (filled circles), compared with results from previous experiments~\cite{ENSDF} (open circles); the latter were shifted 
slightly in $A$ for visual clarity.
}
\label{fig:halflives-results}
\end{center}
\end{figure*}

The deduced $\beta$-decay half-lives  are listed in Table~\ref{tab:halflives-results}, along with their systematic and statistical errors. The statistical error was in most cases significantly larger than the systematic one. For the sake of completeness, 
the half lives obtained from least-square fits of the decay curves for those cases with sufficient statistics are listed as well. The measured half lives for different isotopic chains are show in Fig.~\ref{fig:halflives-results} (filled circles), where only statistical errors are included. The present data agree with previous measurements (open circles) within 1$-$$\sigma$ uncertainty. An exception is $^{88}$Se whose half life reported here is significantly lower than the value found in the literature.

As expected on the basis of the $Q_{\beta}$ evolution, the half-lives systematics follow a decreasing trend with increasing neutron number. 
The discontinuity in the half-live systematics of $_{38}$Sr isotopes, with an abrupt drop at $^{97}$Sr, arises from the presence of high-energy $1/2^{-}$ states in $^{97}$Y$_{58}$ (probably involving $p_{1/2}$ configurations) which absorb most of the $\beta$-decay strength $S_{\beta}$ via allowed transitions from the $1/2^{-}$ ground state of $^{97}$Sr. Such allowed transitions are not possible from the $^{98}$Sr$_{60}$ ground-state $0^{+}$, which decay via forbidden transitions into a daughter's spectrum dominated by negative-parity states. Such states involve the coupling of $\pi p_{1/2}$ with $\nu s_{1/2}$, $\nu d_{3/2}$ and/or $\nu g_{7/2}$, coexisting with negative-parity levels~\cite{Bra89}.

The presence of an $N$=56 sub-shell gap in the region below Zr has motivated numerous studies 
with contradictory outcomes. The semi-magic character of $N$=56 for $Z$=37$-$42 is supported on the basis of the $E2_{1}^{+}$ systematics. As for $^{92}$Kr$_{56}$, the smooth evolution of low-energy states in $^{93}$Kr as compared with $^{95}$Sr was interpreted by Lhersonneau \emph{et al.}~\cite{Lhe01} as a signature of the persistence of the $N$=56 sub-shell for the Kr isotopes. On the other hand, the vanishing of this sub-shell for $^{92}$Kr was suggested by Delahaye \emph{et al.}~\cite{Del06}, 
based on their mass measurement results.
This conclusion is supported by the large quadrupole collectivity deduced from the 
$B(E2; 0^{+} \rightarrow 2^{+})$ measured at REX-ISOLDE~\cite{Muc09} for this isotope. In the light of these results, the unexpectedly high $2_{1}^{+}$ energy measured for $^{88}$Se$_{54}$ by Jones \emph{et al.}~\cite{Jon06} has special significance. 
Such a sudden increase of the $E2_{1}^{+}$ energies for the $N$=54 isotone systematics could be a harbinger of an even higher energy in $^{90}$Se$_{56}$, compatible with a spherical shape. The monopole shift induced by the complete filling of the $d_{5/2}$ neutron orbital for the $N$=56 isotones might push the $\pi f_{5/2}$ level down, and the $\pi p_{3/2}$ level up in energy, creating a $Z$=34 sub-shell. 
Such a scenario would explain the large $E2_{1}^{+}$ measured for $^{88}$Se$_{54}$. In addition, due the complete filling of $\nu d_{5/2}$ for $N$=56, the $^{90}_{34}$Se$_{56}$ nucleus would emerge as a new doubly-magic nucleus. A similar monopole shift has been reported for the proton orbitals in odd-mass $_{29}$Cu isotopes. There, the $5/2^{-}$ ground state assigned to $^{75}$Cu$_{46}$~\cite{Fla09} is an indication of an inversion of the $\pi p_{3/2}$ ground state and
$\pi f_{5/2}$ excited state of $^{69}$Cu$_{40}$, ascribed to the addition of $g_{9/2}$ neutrons.

\begin{figure*}[t!]
\begin{center}
\includegraphics[width=15cm]{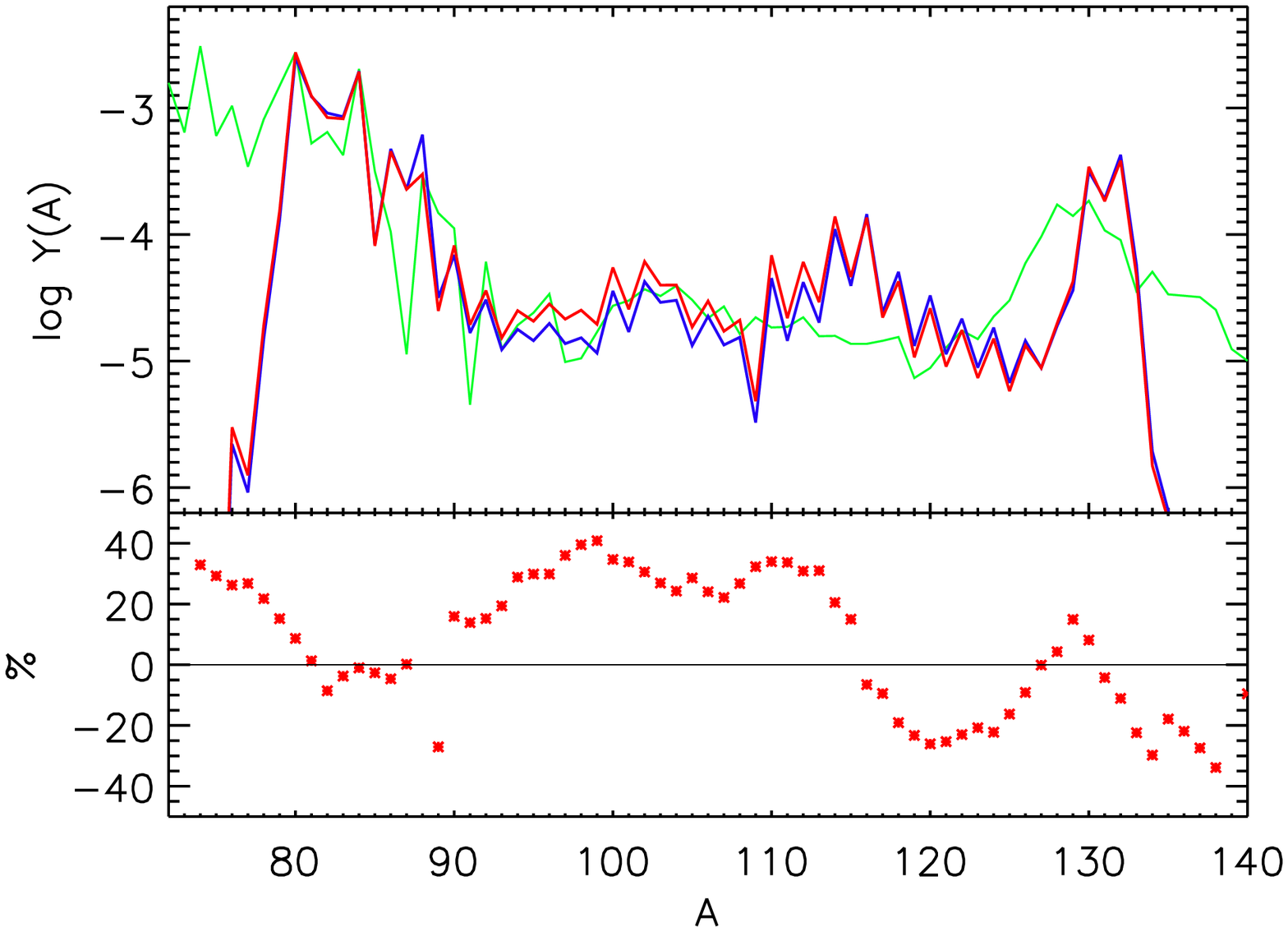}
\caption{(Color online). A comparison of the r-process abundances resulting from a simulation using the FRDM $\beta$-decay lifetime for $^{90}$Se (blue line) and the new experimental lifetime (red line). Outputs of both simulations are shown against the scaled solar abundances (green thin line)~\cite{Sneden2008}.  The bottom panel shows the percentage difference between the two abundance patterns.  The $r$-process simulation uses constant temperature conditions with $T=1.43\times 10^{9}$K and initial neutron density of $N_{n}=5.0\times 10^{22}$/cm$^{3}$. All three curves are normalized to each other at A=80.
}\label{fig:rproc}
\end{center}
\end{figure*}

A doubly-magic $^{90}_{34}$Se$_{56}$ nucleus would have a very different impact on the r-process abundance pattern than a deformed shape. In the top panel of Fig.~\ref{fig:rproc}, the observed solar r-process abundances (green thin solid line) are compared with results obtained from network calculations~\cite{Sur2001} with astrophysical conditions for a weak r-process (temperature $T=1.43\times 10^{9}$K and initial neutron density $N_{n}=5.0\times 10^{22}$/cm$^{3}$). The difference between the solid red and blue lines in the figure is the $\beta$-decay half life of $^{90}$Se. Whereas the former depicts a calculation where half lives and P$_{n}$ values $^{90}$Se were taken  to be spherical from M\"oller's QRPA97 model~\cite{Mol97}, in the latter, the measured half-life of $^{90}$Se was used. The percentage differences between both calculations are shown in linear scale in the bottom panel. It is remarkable how the half life of a hypothetically spherical doubly-magic $^{90}$Se induces variations in the calculated abundances of more than 40$\%$ in the $A$=70$-$140 region.
Moreover, after the neutron freeze-out, the large amount of matter accumulated in this pseudo waiting-point nucleus decays through a series of $\beta$ decays that includes the emission of up to two neutrons. As seen in Fig.~\ref{fig:rproc}, the net result is an overproduction of the stable $^{88}$Sr nucleus. (Note that for $A$=88 the percentage difference is out of scale in the bottom panel of the figure.)

\begin{figure*}[t!]
\begin{center}
\includegraphics[width=7cm]{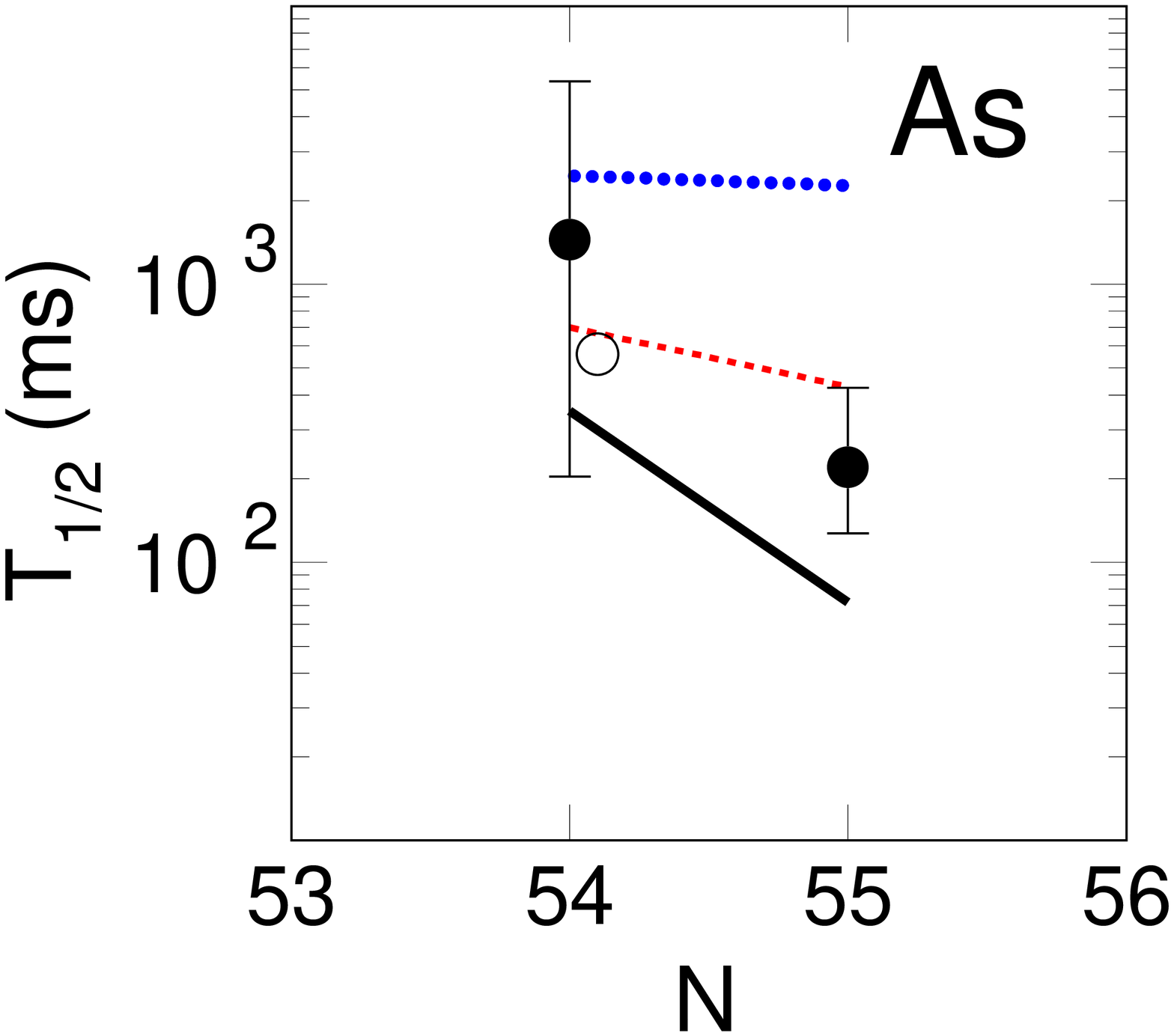}
\includegraphics[width=7cm]{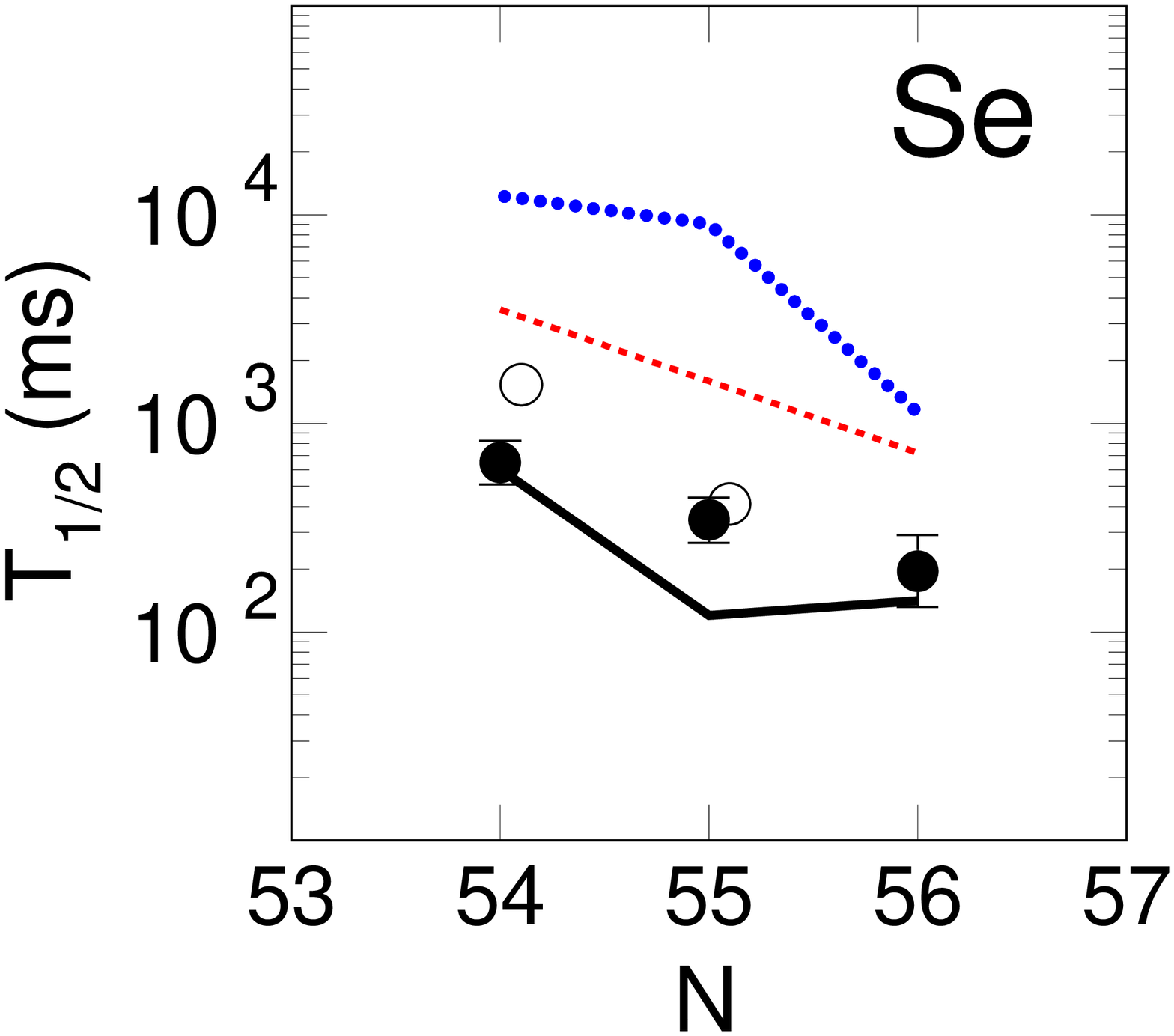}
\caption{(Color online). $\beta$-decay half-lives obtained from the Maximum Likelihood method for As, Se, Br, Kr, Rb, Sr, and Y isotopes (filled dots), compared with results from calculations by M\"oller \emph{et al.} QRPA97~\cite{Mol97} (solid black line) and QRPA03~\cite{Mol03} (dashed red line), both of which use the nuclear deformation parameters from FRDM~\cite{Mol95}. The dotted blue line corresponds to QRPAsph, which was calculated assuming spherical nuclear shapes. The half lives measured in previous experiments~\cite{ENSDF} are shown by open circles. (See text for details.)
}\label{fig:halflives-thresults}
\end{center}
\end{figure*}

The half lives deduced for $_{33}$As and $_{34}$Se isotopes in the present work are compared in Fig.~\ref{fig:halflives-thresults} to values obtained from three different calculations based on the quasi-random phase-approximation (QRPA)~\cite{Kru84,Mol90}. In the first (QRPA97~\cite{Mol97}) and second (QRPA03~\cite{Mol03}) calculations (solid and dashed lines, respectively), the nuclear-deformation parameters were determined using the Finite-range droplet model (FRDM) of M\"{o}ller et al.~\cite{Mol95}. In addition, QRPA03 includes first-forbidden transition rates obtained from the statistical gross theory~\cite{Tak72,Tak73}. The third calculation (QRPAsph) (Fig.~\ref{fig:halflives-thresults}, dotted blue line) was done using QRPA03, assuming spherical nuclear shapes for both, the mother and the daughter isobars.

In the case of $^{87}$As$_{54}$, the three calculations agree with the adopted half life within 1$-$$\sigma$ error. However, the more precise half life for this nucleus from the ENSDF database~\cite{ENSDF} (obtained by statistical evaluation among three independent experiments) and the new half life for $^{88}$As are only compatible with QRPA03.



Similar results were obtained for the half lives of the $_{34}$Se isotopes. The half lives calculated with QRPAsph are too high presumably due to the dominance of forbidden transitions between levels of opposite parity, involving $f_{5/2}$, $p_{3/2}$, $p_{1/2}$ protons, and $d_{5/2}$, $g_{7/2}$, $s_{1/2}$, $d_{3/2}$ neutrons.
QRPA97 provides the best agreement with the new measured half life of $^{90}$Se, implying that this nucleus is not doubly magic but in fact, deformed. According to Ref.~\cite{Mol97}, $^{90}$Br is the first neutron-rich Br isotope in which $\pi g_{9/2}$ defines the ground state via coupling between its $1/2^{+}$ state and the $3/2^{+}$ neutron state from $\nu d_{5/2}$. In this context, the new half life of $^{90}$Se obtained here rules out the possibility of a spherical waiting-point that would contribute to the overproduction of stable Sr, Y, and Zr isotopes observed in some metal-poor stars~\cite{Hon06}.

\section{Summary}
The $\beta$-decay half lives of $_{33}$As, $_{34}$Se, $_{35}$Br, $_{36}$Kr, $_{37}$Rb, $_{38}$Sr, and $_{39}$Y isotopes around $N$=56 were measured at the National Superconducting Cyclotron Laboratory. New half lives are reported  for $^{88}$As$_{55}$ and $^{90}$Se$_{56}$ for the first time.

The $\beta$ decays from the nuclei of interest were observed with the Beta Counting System. The half-lives were deduced using the Maximum-Likelihood method and, where statistics were sufficient, least-squares fits of decay curves. Agreement between both analyzes brings confidence to the results. For most of the cases, the data agree with previous measurements. 

The structure of $N$=56 isotones in this region was investigated by comparing the measured $T_{1/2}$ values with QRPA calculations. The deformed intruder states from $\pi g_{9/2}$ and $\nu h_{11/2}$ play a crucial role creating allowed transitions that compensate the slow forbidden transitions dominating the $\beta$ decay of a spherical $^{90}$Se.

No evidence was found for a $N$=56 sub-shell closure in $^{90}$Se$_{56}$, let alone the emergence of a new $Z$=34 semi-magic number due to the monopole interaction between $\pi f_{5/2}$ and $\nu d_{5/2}$ orbitals. Nevertheless, given the gross nature of $\beta$-decay $T_{1/2}$ as a nuclear-structure probe, more precise spectroscopic studies of $^{90}$Se$_{56}$ are needed.

\vspace{1cm}

\begin{acknowledgments}
The authors wish to thank the NSCL operations staff and R.~Fox for their help in the preparation and running of the experiment.

This work was supported in part by the Joint Institute for Nuclear Astrophysics (JINA) under NSF Grant PHY-02-16783, PHY 08-22648, the Institute of Nuclear Structure and Astrophysics under NSF grant PHY-07-58100, and the National Superconducting Cyclotron Laboratory (NSCL) under NSF Grant PHY-01-10253 and PHY-06-06007.
\end{acknowledgments}

\end{document}